# A Hands-On Guide

## to

## Shear Force Mixing of Single-Walled Carbon Nanotubes with Conjugated Polymers

Sebastian Lindenthal, Simon Settele, Joshua Hellmann, Klaus Schmitt,

and Jana Zaumseil

Institute for Physical Chemistry, Universität Heidelberg, D-69120 Heidelberg, Germany

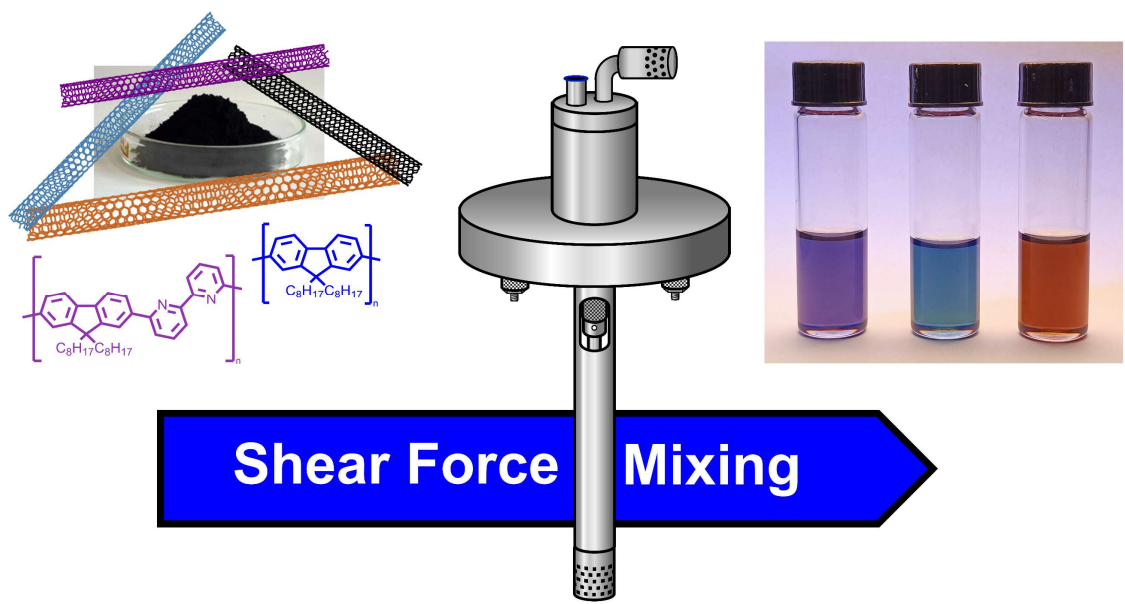

**Version 1.0 / 12.11.2023**

# ABSTRACT

This guide provides a detailed step-by-step procedure for the dispersion of (6,5) single-walled carbon nanotubes by shear force mixing with the conjugated polymer PFO-BPy in organic solvents. All processes presented here were developed in the Zaumseil group at Heidelberg University since 2015 and represent best practices to the best of our knowledge. In addition to the detailed instructions, we discuss potential pitfalls and problems, that we have encountered over eight years of operation and show how to solve them. This also includes a detailed description of how to maintain and service a shear force mixer to ensure long operation lifetime. Finally, we show how to expand our process to the dispersion other nanotube chiralities in electronic-grade quality and how to treat dispersions for subsequent processing (*e.g.*, thin film deposition or functionalization).

**If you use this guide please cite it as:**

https://doi.org/10.11588/heidok.00033977

**as well as the original publication:**

Graf, A.; Zakharko, Y.; Schießl, S. P.; Backes, C.; Pfohl, M.; Flavel, B. S.; Zaumseil, J., Large scale, selective dispersion of long single-walled carbon nanotubes with high photoluminescence quantum yield by shear force mixing. *Carbon* **2016**, *105*, 593-599.

**Acknowledgements**: We thank Arko Graf and Jan Lüttgens for the initial development and optimization of the Shear Force Mixing process. Special thanks go to the dedicated undergraduate students who ran the Shear Force Mixing process within our group over the past years. Our deepest gratitude is reserved for our mechanical workshop team, led by Klaus Schmitt, whose support was indispensable whenever challenges arose with our Shear Force Mixers.

**Funding:** This work was financially supported by the European Research Council (ERC) under the European Union's Seventh Framework Programme (FP/2007–2013) (Grant Agreement No. 306298 "EN-LUMINATE") and under the European Union's Horizon 2020 research and innovation programme (Grant Agreement No. 817494 "TRIFECTs").

# CONTENTS

# INTRODUCTION TO SWNT DISPERSION

The optical and electronic properties (metallic or semiconducting, bandgap) of single-wall carbon nanotubes (SWNTs) are determined by their chiral indices (n,m) (see **Figure 1a**). Commercial growth methods such as CoMoCAT®, HiPco® or the Plasma Torch process always yield a mixture of different nanotube chiralities with different diameter distributions and metallic nanotube content.[1,2] To integrate SWNTs in useful devices, methods to form stable dispersions and select desired chiralities are required. De-bundling of the as-synthesized SWNTs is achieved by applying sufficient mechanical force *via* sonication or shear force mixing in a solvent followed by stabilization through surfactants or polymers. Stable dispersions in water have been achieved by addition of surfactants such as sodium cholate or sodium dodecyl sulfate[3,4] or single-stranded DNA and RNA.[5,6] These dispersants are, however, mostly unspecific toward nanotube chirality, which makes sorting of the SWNTs *via* methods such as aqueous two-phase extraction[7,8], density gradient ultracentrifugation[9] or column chromatography[10] necessary.

Selective dispersions of nanotubes with suitable conjugated polymers in organic solvents provides a direct route to certain nanotube types (semiconducting) and even single chiralities. Fluorene-based polymers and copolymers show especially high selectivities. **Figure 1b** shows the preferentially selected (n,m) chiralities for a few combinations of polyfluorene polymer and SWNT raw material.[11-16] In addition, polymer-related (molecular weight, concentration, polymer to nanotube ratio),[14,17,18] solvent-related (polarity, viscosity)[14,19,20] and process-related (sonication time, temperature)[14,21-23] parameters have an impact on the yield and selectivity of the dispersion process. For more details, we refer the reader to two review articles by Samanta *et al.*[24] and Fong *et al.*[25] which discuss the influence of these parameters extensively.

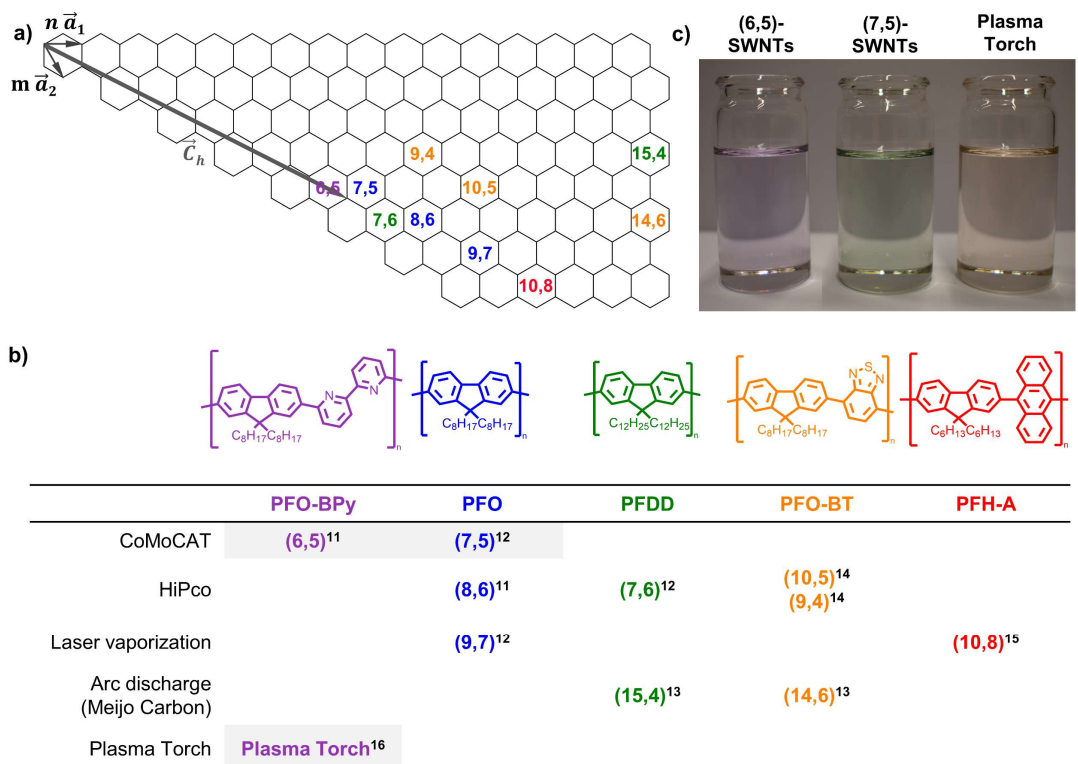

**Figure 1. a)** Schematic depiction of a graphene sheet with the chiral vector $\vec{C}_h$ for a (6,5) SWNT. Chiral indices of nanotube chiralities that can be selected by polymer-wrapping are shown on the honeycomb lattice. **b)** Structures of highly-selective wrapping polymers and an overview of their selectivity for dispersion of different starting materials in toluene. In this work we provide a protocol for shear force mixing of the grey-shaded chiralities. **c)** Photographs of purified dispersions of (6,5), (7,5) and semiconducting Plasma Torch SWNTs in toluene obtained by Shear Force Mixing with polyfluorene polymers.

## SHEAR FORCE MIXING as an EXFOLIATION METHOD

The quality of a nanotube dispersion (*i.e.*, the defectiveness and length of the contained nanotubes) is strongly influenced by the applied dispersion method. Structural defects and open tube ends will act as quenching sites for excitons (strongly bound electron-hole pairs that are highly mobile on carbon nanotubes), thus decreasing their emission efficiency.[26] Additionally, defects and inter-tube junctions in SWNT thin film devices, can act as charge-carrier scattering sites and will reduce charge-carrier mobilities.[27, 28] Both effects significantly influence the performance of SWNTs in photonic, electronic, and optoelectronic devices, which makes long and defect-free SWNTs highly desirable.

Sonication with a bath or tip sonicator damages and shortens nanotubes due to highly localized friction and pressure, created by sonication-induced cavitation.[29, 30] In contrast to that, Shear Force Mixing, provides energy for exfoliation continuously without locally creating highly turbulent and damaging regions. A basic explanation of the working principle of a Shear Force Mixer in the context of nanotube exfoliation is given in **Figure S1**.

Shear force mixing was previously shown to be a mild dispersion method for 2D materials (*e.g.*, graphene, $MoS_2$)[31, 32] and for SWNTs in aqueous dispersions[33] that enables larger flakes and longer nanotubes than sonication methods. Graf *et al.* showed that using shear force mixing of CoMoCAT® nanotubes with a polyfluorene copolymer (PFO-BPy) in toluene results in long nanotubes (> 1 µm) with low defect densities.[34] The average values for nanotube length, Raman $D/G^+$-ratios (as a metric for defect density) and photoluminescence quantum yields (PLQY) for different exfoliation methods are shown in **Table 1**. While Graf *et al.* reported a comparatively high $D/G^+$-ratio of 0.10 for shear force mixing in their original paper, we nowadays observe values of <0.05 for dispersions obtained by shear force mixing.

**Table 1**. Average lengths, $D/G^+$-ratios and PLQY for PFO-BPy wrapped (6,5) SWNTs exfoliated by Shear Force Mixing, Bath Sonication, Tip Sonication and (6,5) SWNTs sorted by Gel Chromatography + Density Gradient Ultracentrifugation (GC-DGU). All values were taken from Graf *et al.* [34]

|  | SFM | Bath Sonication | Tip Sonication | GC-DGU |
| --- | --- | --- | --- | --- |
| **Length (µm) Mean/Median** | 1.82/1.55 | 1.12/1.01 | 0.61/0.50 | 0.62/0.58 |
| **$D/G^+$-ratio** | 0.10 | 0.11 | 0.08 | 0.10 |
| **PLQY (%)** | 2.3 | 1.8 | 1.3 | 0.2 |

In addition to the superior nanotube quality, shear force mixing offers a scalable approach to exfoliate large quantities of nanotubes in one batch. Exfoliation of SWNTs and other nanomaterials by Shear Force Mixing is possible in large volumes up to 300 L.[31]

In the following we provide a step-by-step protocol for the dispersion of long, defect-free (6,5) and (7,5) SWNTs from CoMoCAT® raw material and purely semiconducting Plasma Torch-SWNTs from Plasma Torch (Raymor Nanotech) raw material by shear force mixing (resulting dispersions shown in **Figure 1c**).

# WORKFLOW OVERVIEW

The workflow for Shear Force Mixing is shown in **Figure 2**. In a first step, SWNTs are weighed and dried (**steps 1-9**), as air humidity and moisture in the material significantly reduce the dispersion yield. The dried raw material is then dispersed in toluene with a suitable wrapping polymer (**steps 10-21**). The exfoliation is followed by centrifugation (**steps 22-30**) to remove unexfoliated material. Since a single SFM run usually does not disperse all (6,5) SWNTs that are contained in the raw material, the pellet from the centrifugation can be recycled (**steps 36-41**) to increase the yield. The supernatant after centrifugation is the final (6,5) SWNT dispersion, which is characterized by absorption and Raman spectroscopy (**steps 42-43**) to determine the yield and the quality of the dispersion.

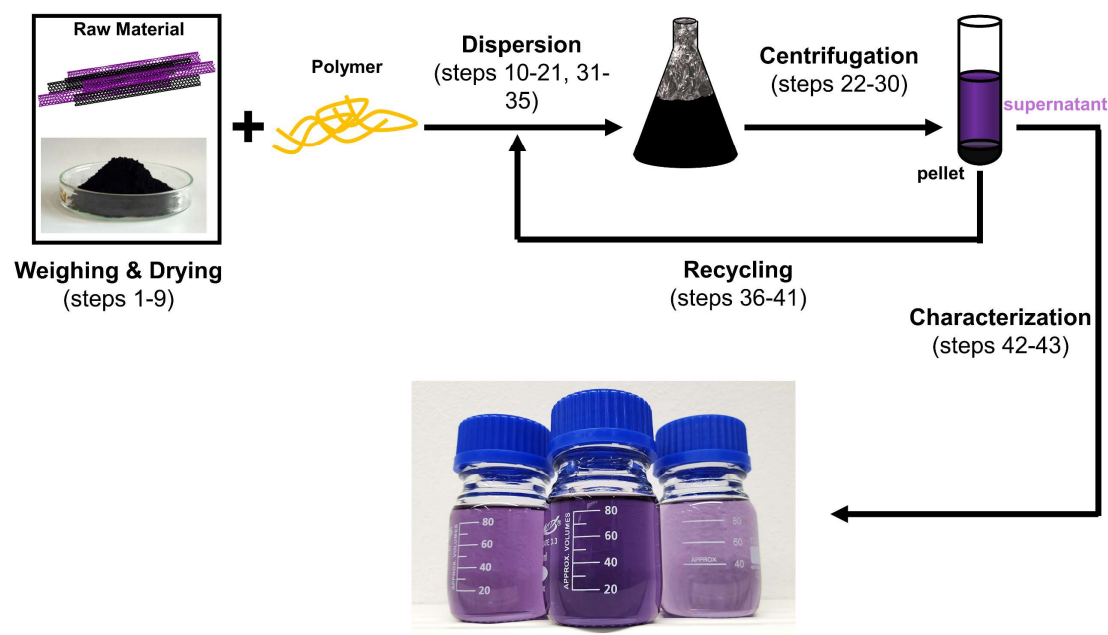

**Figure 2.** Workflow of the Shear Force Mixing process. Nanotube raw material is weighed and dried. Shear Force Mixing with a suitable wrapping-polymer and toluene yields a dispersion that must be purified by centrifugation. The resulting supernatant is characterized while the pellet can be recycled several times to maximize the yield.

In the following we present a step-by-step protocol for the monochiral dispersion of (6,5) SWNTs *via* Shear Force Mixing. While the dispersion protocol is similar for other chiralities, there are some important differences. The experimental variations are explained in the Supporting Information (**Appendix I** for (7,5) and **Appendix II** for Plasma Torch SWNTs).

In case excess polymer is problematic (*e.g.*, for functionalization or thin film deposition experiments), you can remove the polymer either by filtration or ultracentrifugation. We added an explanation for both processes in **Appendix III**.

## EQUIPMENT

Here we list the materials, consumables and the equipment that is necessary for the dispersion of (6,5) SWNTs with PFO-BPy in toluene. For other chiralities see the Supporting Information. Specific companies or brands are named to describe our process as detailed as possible and do not represent any endorsement and recommendation.

**Required Raw Materials / Polymers / Solvents**

- **CoMoCAT® raw material** from CHASM (SG65i, *e.g.*, Sigma-Aldrich, 773735, (6,5) chirality, ≥95% carbon basis (≥95% as carbon nanotubes), 0.78 nm average diameter)
- **Poly[(9,9-dioctylfluorenyl-2,7-diyl)-alt-(6,6′-[2,2′-bipyridine])]** (American Dye Source, PFO-BPy, ADS153UV, $M_w$ ~ 34 000 g/mol, note that the molecular weight range is critical and always should be between 30 000 and 40 000 g/mol)
- **Toluene** (purity >99.7 %)
- **Isopropanol**

**Required Equipment for Nanotube Weighing and Drying (steps 1-9)**

- **High precision scale (**e.g. A&D Weighing HR-250AZ Galaxy Analytical Balance**)**
- **Polyimide-tape** (3M™, 5419 high temperature polyimide tape)
- **PTFE filter** (Millipore, Omnipore Membrane Filter JHWP02500, hydrophilic PTFE Filter, 25 mm diameter)
- **Snap-cap vials** (VWR, 15 mL, ND22 opening)
- **Spatula**
- **Tweezers**
- **Wipes** (Kimtech Science, Precision Wipes)
- **Hotplate**

**Personal protective equipment (PPE):**

- **Safety glasses and lab coat** (obviously…)
- **Particle mask** (Dräger, model type: X-plore 3300, equipped with bayonet filter type: A1B1E1K1 Hg P3 R D) or similar
- **Nitrile gloves** (Ansell, Touch N Tuff® 92-600, Length 240 mm)
- **Long-sleeve gloves** (Hygostar, LDPE gloves Softline long)

**Required Equipment for Dispersion (steps 10-21)**

- **Shear Force Mixer** (L2/Air, Silverson Machines Ltd.) depicted in **Figure 3**.

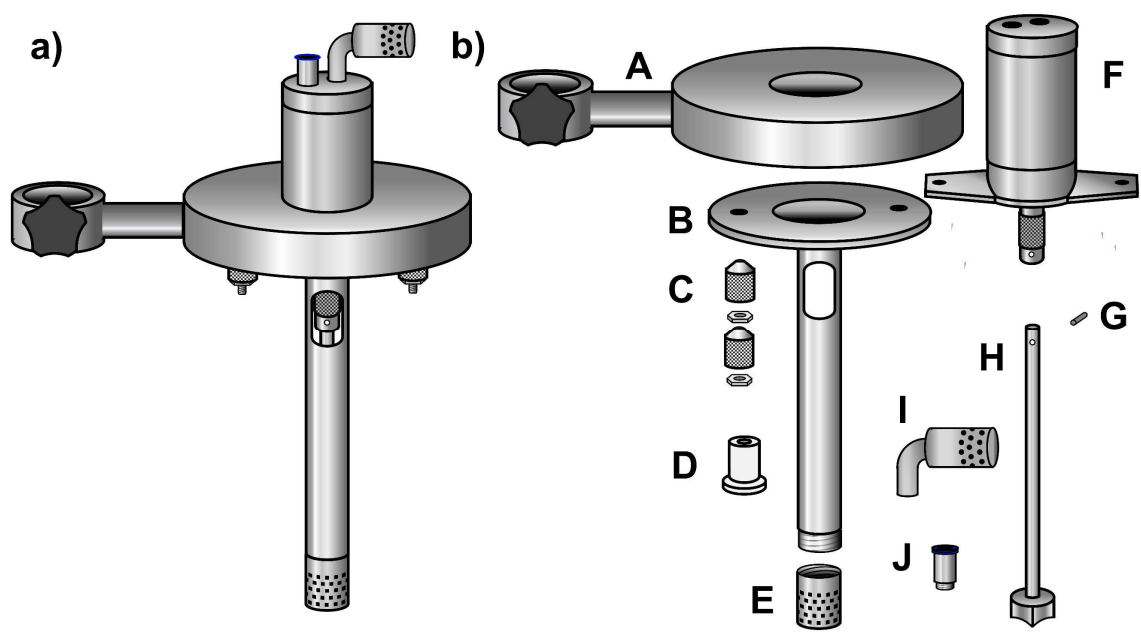

**Figure 3. a)** Assembled shear force mixer **b)** Explosion drawing of shear force mixer parts with a detailed description given in Table 2.

**Table 2.** List of the shear force mixer parts.

| | |
|---|---|
| **A** | Metal block for attachment of F and B with connection to lab stand |
| **B** | Frame for detachable shear heads (in this paper the combination of B and E will be referred to as stator) |
| **C** | Conical screws and nuts (for connection between A and B) |
| **D** | ¾" PTFE Bush (CAB6E10/1) |
| **E** | **¾" square hole high shear screen** (here we call the combination of B and E stator) |
| **F** | Compressed air motor (**Atlas Copco, LZB 22RL A220-11**, 0.34 HP (0.25 kW) lubrication free stainless-steel air motor (9600 RPM at 5.3 L/s (11.2 cfm))) |
| **G** | Aglet for connection of F and H |
| **H** | Drive shaft with ¾" rotor |
| **I** | Exhaust for compressed air motor |
| **J** | Push-in fitting for connection to PVC hose (Landefeld, 3/8" IQS straight push-in fitting, diameter 12 mm) |

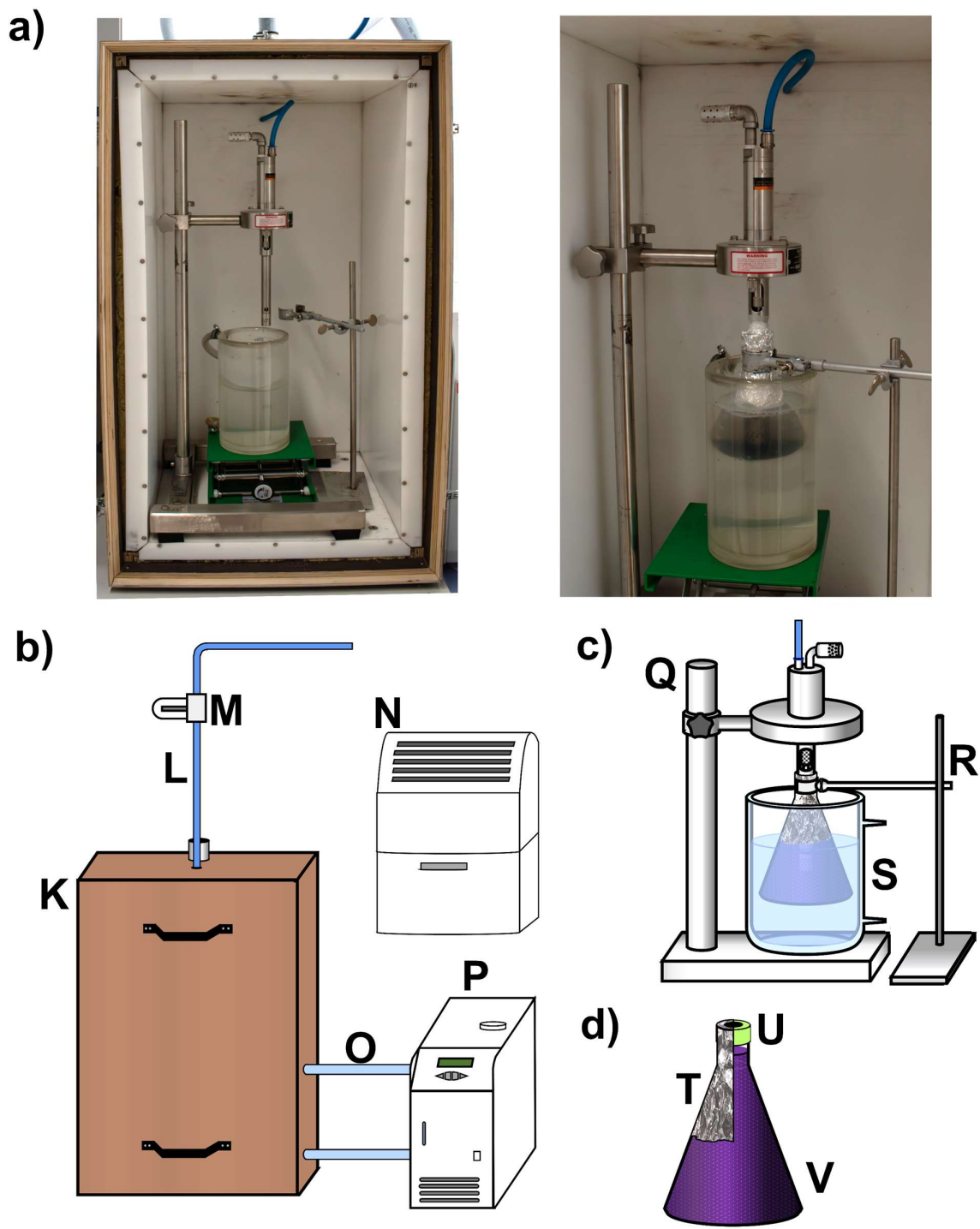

**Figure 4. a)** Pictures of the Sound proof box that contains the Shear Force Mixer. Left panel only shows the Shear Force Mixer. Right Panel shows how the setup looks when an Erlenmeyer flask is attached to it. **b-d)** Schematic depiction of additional equipment needed for dispersion of SWNTs with Shear Force Mixing.

**Table 3.** Detailed list of required accessories for the Shear Force Mixing process

| K | Soundproof box containing the SFM setup (Homebuilt, Explanation and technical drawing can be found in **Appendix IV** of the **Supporting Information**) |
|---|---|
| L | Polycarbonate hose (Landefeld, 10.2x8 mm diameter, outside/inside) with connection to pressurized air source with 5 bar pressure |
| M | Air filter (Landefeld, multifix series 0 - type F014, 1000 l/min) |
| N | Dehumidifier (Remko, model type: ETF320) |
| O | PVC hoses for cooling water (Landefeld, 16x10 mm diameter, outside/inside) |
| P | Chiller (Julabo GmbH, model type: F250) |
| Q | Stainless steel metal lab stand for the shear force mixer |
| R | Lab stand for attachment of Erlenmeyer flask |
| S | Tempering beaker (KGW-Isotherm GmbH, model type: T 2000) |
| T | Aluminum foil and parafilm |
| U | Splash protection (Homebuilt, made from PP lid of centrifuge tubes) |
| V | Erlenmeyer flask (VWR, borosilicate glass, NS29/32, 250 mL) |

**Required Equipment for Purification (steps 22-30)**

- **Ultracentrifuge** (e.g., Beckmann Coulter, Avanti J-26S XP equipped with a JA25.50 fixed-angle rotor, the centrifuge needs to reach RCF values of 60 000 $g$)
- **Centrifuge tubes** (Beckmann, **Polypropylene** Centrifuge Tubes with PE Caps, 50 mL)
- **Eppendorf pipette** (10 mL)
- **Pipette tips**
- **Syringes** (BD, Discardit II, 20 mL)
- **Syringe Filters** (Whatman, Puradisc 13, PTFE filters 5 μm pore size, 13 mm diameter)

**Required Equipment for Cleaning of the SFM (steps 31-35)**

- **Test tube brush**
- **Measuring cylinder** (must be big enough so that the stator can fit inside)
- **Ultrasonication bath** (here, Branson 2510)

**Required Equipment for Characterization (steps 42-43)**

- **Absorption spectrometer with detection into the nIR-region** (here, Cary 6000i UV-Vis-NIR absorption spectrometer, Varian, Inc.)
- **Cuvettes for absorption spectroscopy (1 cm pathlength)**
- **Raman spectrometer with suitable excitation wavelengths** (here, Renishaw inVia confocal Raman microscope in backscattering configuration equipped with a 50× long working distance objective (N.A. 0.5, Olympus) and with excitation wavelengths of 532 nm, 633 nm and 785 nm)
- **Glass coverslips, aluminum foil**
- **Hotplate**

# PROTOCOL for DISPERSION of (6,5) SWNTs

Here we describe the standard shear force mixing process for the dispersion of 125 mg of CoMoCAT® raw material (SG65i) with 125 mg of PFO-BPy in 250 mL toluene to obtain a nearly monochiral (6,5) SWNT dispersion. You can scale down the process to a volume of 125 mL by simply halving the amount of raw materials and using a smaller Erlenmeyer flask. The process stays the same.

**Nanotube weighing** ● **TIMING 1 h (+ at least 10 h drying time)**

▲**CRITICAL** The CoMoCAT® raw material should be stored in a dry environment, such as a glovebox or dry box, to minimize the influence of humidity on the dispersion process.
▲**CRITICAL** Use dehumidifiers in the labs (if not air-conditioned and humidity-controlled) where you handle and disperse nanotubes, as high humidity can decrease the dispersion yield and quality.
**! CAUTION** Dry nanotube raw material can be electrostatically charged and forms particle aerosols that pose a respiratory hazard. Always ensure to work in a fume hood and avoid spills. Clean the equipment that was in touch with nanotube raw material (spatulas, tweezers, fume hood) with IPA-wetted Kimtech wipes until no more black particles can be seen on the wipes.

1  Clean the fume hood in which you will weigh the nanotubes thoroughly by wiping it down with a Kimtech wipe and IPA.

2  Line the fume hood with Kimtech wipes and wet them with IPA. Ensure that the wipes stay wet as long as you work with the nanotubes, so spilled nanotube raw material sticks to the wet wipes.
   **! CAUTION** Ensure to remove any ignition sources (*e.g.* hotplates) from the fume hood where you are working.

3  Cut six stripes of Kapton tape (length ~4-5 cm) and stick them to a surface but ensure they can be removed from it later. You only need four stripes of tape, two are just for backup.

**4**  Place a snap-cap vial, the container with CoMoCAT® raw material and a long-sleeve glove (will act as a trash can) on the IPA-wetted Kimtech wipes. Place a spatula, tweezers and a PTFE filter on a dry spot in the fume hood.

**5**  Put on a lab coat, the particle mask and a pair of nitrile gloves. Wear a pair of long-sleeve gloves over the nitrile gloves. Put on a last pair of nitrile gloves on top of the long-sleeve gloves.

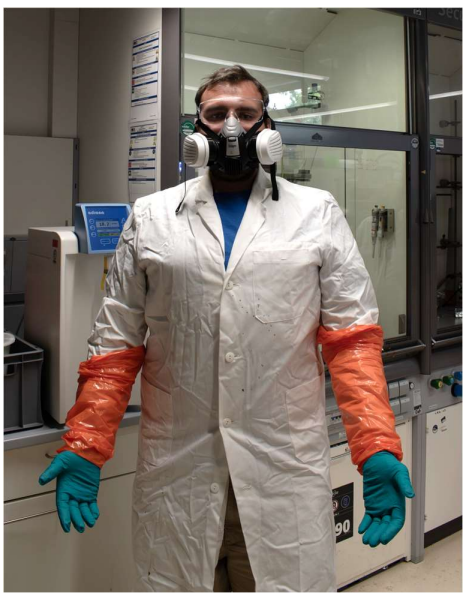

**Figure 5.** Personal protective equipment for weighing of nanotubes. Ensure that the long sleeve gloves extend up to your elbows, so that in case of nanotube spills only the gloves and not the lab coat will be contaminated.

**6**  Tare (zero) the snap-cap vial and weigh in 125 mg of nanotubes with the spatula.
▲CRITICAL The bottom of the snap-cap vial will be wet from the IPA-wet Kimtech wipes during the weighing process. Thus, the mass will change (loss of 1-2 mg) when the IPA starts evaporating on the scale.
▲CRITICAL Do not try to weigh the nanotubes directly on the scale. Instead place the snap-cap vial on the Kimtech wipes when you transfer them from the container. Hold the container with the nanotubes as close as possible over the snap-cap vial to minimize the number of nanotubes that fly off the spatula. If the raw material exhibits a high electrostatic charging behaviour try to use an anti-static gun on the nanotube raw material.
▲CRITICAL Try to weigh in 125 mg, but anything from 120-130 mg is fine. Do not start to remove nanotubes from the vial again, as it will create more nanoparticle aerosol.

| 7 | Use tweezers to place the PTFE filter membrane on the snap-cap vial (**Figure 6a**) and adhere the PTFE filter to the vial using 4 stripes of Kapton tape leaving a quadratic window uncovered so that water vapour can still escape (**Figure 6b**). |

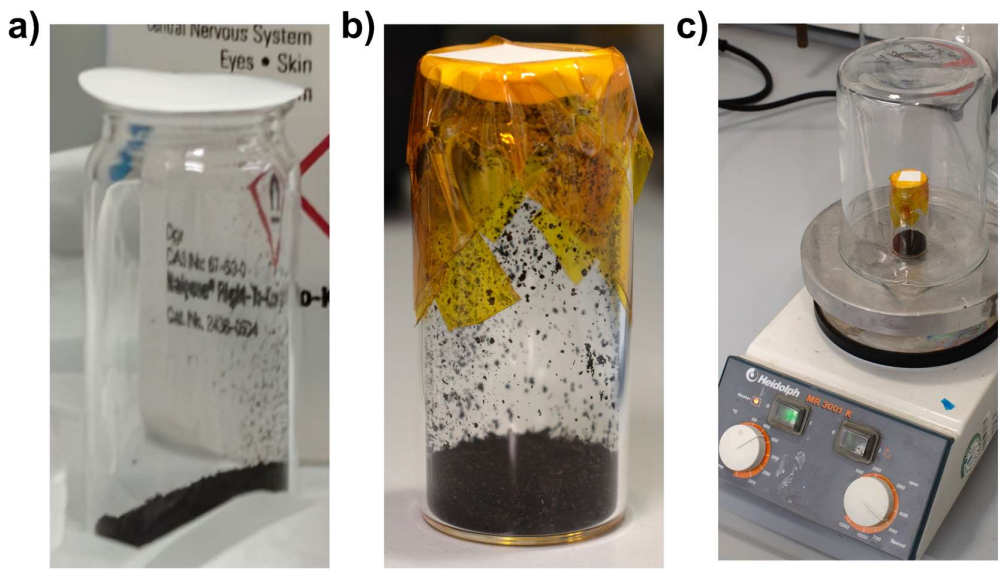

**Figure 6. a)** Vial with 130 mg CoMoCAT® raw material and loose PTFE filter. **b)** Vial sealed with PTFE filter and Kapton tape. **c)** Vial placed on hotplate for drying. The beaker is placed on top to prevent people from accidentally knocking over the vial with SWNTs.

| 8 | Place the vial on a hotplate and heat it to 130 °C for at least overnight (**Figure 6c**). ▲CRITICAL Drying of the raw material is **<u>not optional!</u>** Not dried material will lead to a significant loss in yield. ■ PAUSE POINT The CoMoCAT® raw material can be dried for up to 3 days without causing any damage to the SWNTs. |

| 9 | Thoroughly wipe down the fume hood with IPA and Kimtech wipes until you see no more black particles. Dirty Kimtech wipes, as well as your outer nitrile gloves and the long-sleeve gloves should go into the long-sleeve glove acting as a trash can. When you are finished press out residual air and tie a knot into the long-sleeve glove. This will seal the waste that is contaminated with nanoparticles (→ solid waste container). |

**Setting up the SFM**                     ● **TIMING** 1 h (+ at least 72 h dispersion time)

▲CRITICAL The polymer should be stored in a dry environment, such as a glovebox or dry box, to minimize the influence of humidity on the dispersion process.

▲CRITICAL Use dehumidifiers in rooms where you handle and disperse nanotubes, as high humidity can decrease the dispersion yield and quality.

10  In an Erlenmeyer flask, prepare a solution of 125 mg of PFO-BPy in ~200 mL of toluene. If the polymer does not fully dissolve upon stirring, use sonication to break up bigger chunks and heat the solution to 60 °C.
    ▲CRITICAL The polymer solution must cool down to room temperature again before you add the dried nanotubes.

11  Remove the nanotubes from the hotplate, so they can cool down to room temperature.

12  Put on your personal protective equipment (safety glasses, lab coat, gloves, particle mask).

13  Place a Kimtech wipe in the fume hood and wet it with IPA. Above this wipe, remove the PTFE filter membrane from the snap-cap vial and pour the dried nanotube material into the polymer solution. Rinse out the snap-cap vial with toluene from a squeeze bottle and pour it to the polymer solution. Rinse down any nanotube raw material that sticks to the inner wall of the Erlenmeyer flask and has not yet been in contact with the polymer solution.
    ! CAUTION Dry nanotube raw material can be electrostatically charged and forms particle aerosols that pose a respiratory hazard. If any nanotube spills happen, thoroughly clean the fume hood with IPA and wipes.

14  Assemble the Shear Force Mixer as shown in **Figure 7**.
    ▲CRITICAL After assembling the shear force mixer, it must be levelled using a spirit level.
    ! CAUTION Ensure that the shear force mixer is running in a well-ventilated space. In our case, the soundproof box is connected to an exhaust. You can also place the SFM in a fume hood.

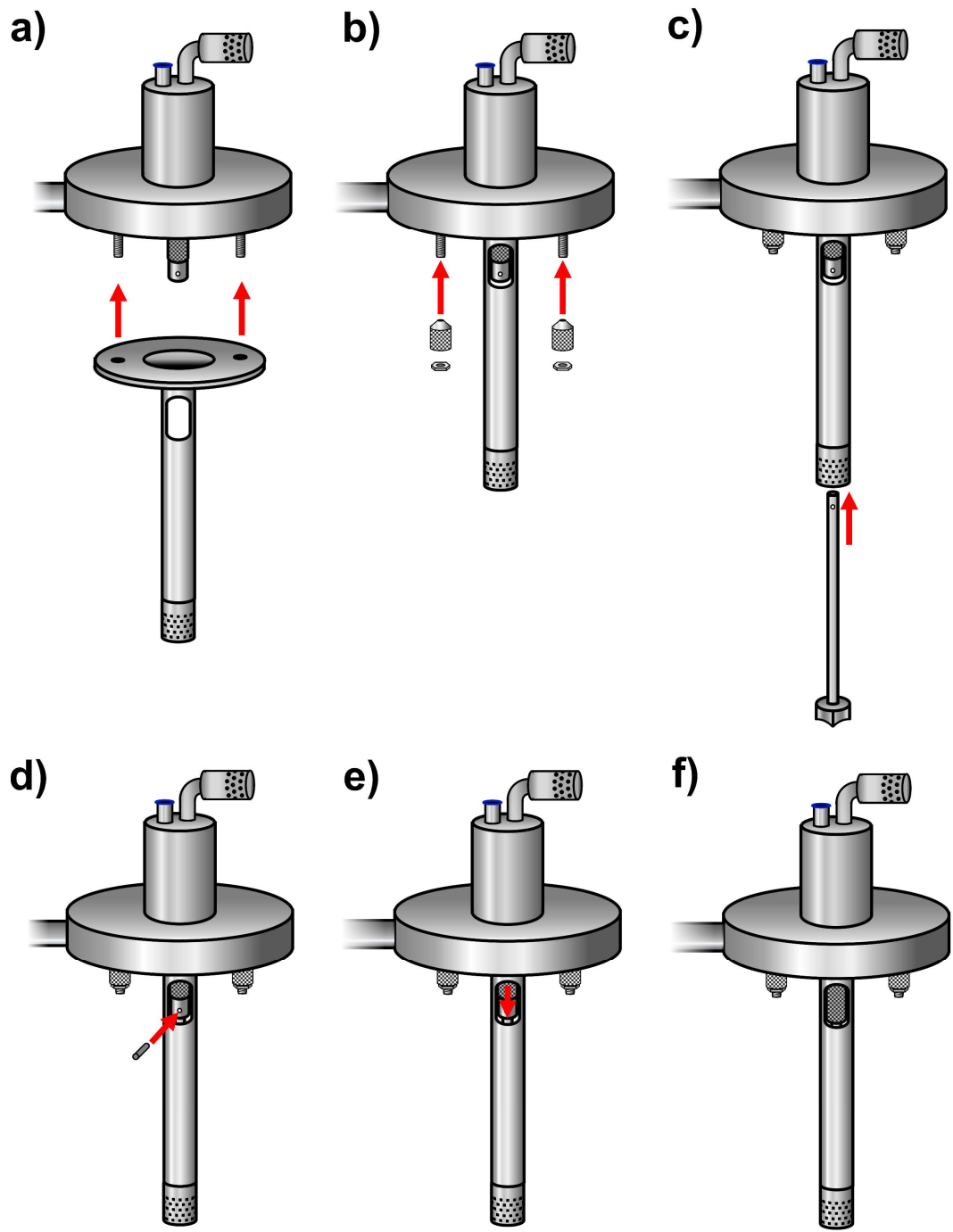

**Figure 7.** Assembly of the SFM. **a)** Insert the stator into the bolts that come out of the metal block. **b)** Use the conical screws and nuts to fixate it. **c)** Insert the rotor into the stator. Align the hole of the rotor with the hole of the motor unit. **d)** Fixate the rotor by inserting the metal aglet into the aligned holes. **e)** Pull the metal protection cap down. **f)** Assembled SFM.

**15**   Attach the Erlenmeyer flask to the lab stand. Make sure that the rotor/stator of the SFM is centered in the middle of the flask and resides ~2 cm above the bottom of the Erlenmeyer flask (**Figure 8**).

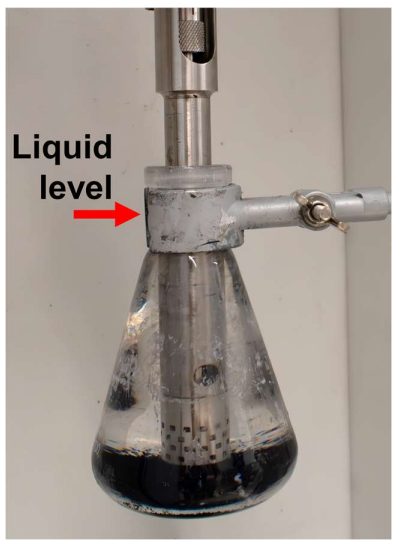

**Figure 8**. The stator is centred in the Erlenmeyer flask and immersed into the solvent. It resides approximately 2 cm above the bottom of the Erlenmeyer flask. The level of the solvent is at around half the height of the bottleneck.

**16**   Fill up the flask with toluene until the level of the liquid reaches half the height of the bottleneck (indicated by an arrow in **Figure 8**).

**17**   Remove the flask again without changing the height of the clamp on the lab stand. Then attach the splash protection (lid of centrifuge tube with hole in it) to the Erlenmeyer flask using aluminum foil and parafilm (see **Figure 9**). The aluminum foil should extend above the splash protection.

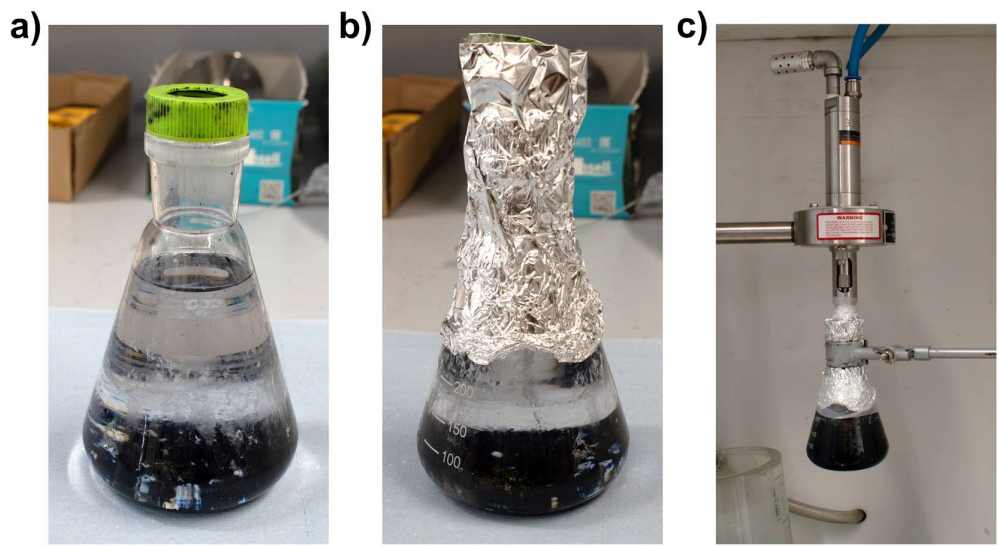

**Figure 9. a)** Erlenmeyer flask with dispersion with splash protection. **b)** Erlenmeyer flask with splash protection, fixated by aluminum foil. **c)** Erlenmeyer flask attached to the SFM

18    Attach the Erlenmeyer flask to the lab stand with the rotor/stator already inserted in the flask. Seal the extension of the aluminum foil around the SFM frame and wrap the connection between them with parafilm (**Figure 9c**)

**! CAUTION** Aluminum foil and parafilm must not cover the opening at the top of the stator frame. If loose ends are pulled into the rotor unit, they will damage the SFM.

19    Without closing the soundproof box, slowly open the valve to the pressurized air. We use an overpressure of 4 bar, which enables the motor to run at its maximum speed of 10230 rpm. While opening the valve, closely monitor the movement of the rotor (does it start spinning?) and listen to the sounds the motor makes. The latter should give off a whining noise with a uniform pitch that rises when you increase the air pressure. If everything looks and sounds good, close the valve to the pressurized air again.

**! CAUTION** Make sure that the protective lid covers the aglet that connects rotor and motor unit. If this is not the case, the aglet will fly out of the setup, becoming a high-speed projectile.

**! CAUTION** If the SFM makes loud noises or the whining noise is not uniform, immediately close the valve to the pressurized air.

**? TROUBLESHOOTING**

20. Bring the cooling beaker up to a position in which the water covers around 2/3 of the height of the Erlenmeyer flask. Turn on the chiller, set the temperature to 20 °C and cover the beaker with aluminum foil to minimize evaporation of cooling water.

    ▲CRITICAL As shown by Gomulya *et al.*, temperature is an important parameter that affects dispersion quality and yield, if not controlled precisely.[23] High temperatures can cause gelification of the mixture and lead to unpurifiable dispersion results. Worst case scenario: All of the solvent evaporated and the rotor will be destroyed! Thus, always make sure to turn on your chiller and check the water level in the cooling beaker every two days during the SFM process.

21. Close the soundproof box and fully open the valve to the pressurized air. Let the shear force mixer run for the desired time.

    ▲CRITICAL ■ PAUSE POINT As shown by Graf *et al.* the dispersion time impacts the yield (see **Figure S2**).[34] We usually let our SFM run for 72-96 h. Further increasing the dispersion time only leads to a slight increase in yield.

**Purification of the stock dispersion**  ● TIMING 3 h

▲CRITICAL Use dehumidifiers in rooms where you handle and disperse nanotubes, as high humidity can decrease the dispersion yield and quality.

▲CRITICAL As soon as the dispersion process is stopped, work as quickly as possible. Even well dispersed (6,5) SWNTs will start aggregating after a few hours if the dispersion is not mixed anymore which leads to sedimentation during the centrifugation process.

22. Close the valve to the pressurized air to stop the shear force mixer. Remove the Erlenmeyer flask from the shear force mixer.

23. Distribute the black dispersion from the Erlenmeyer flask equally into the centrifuge tubes and place them in the rotor of the centrifuge.

    ▲CRITICAL As the centrifuge tubes are rather expensive we usually use them several times.

> **! CAUTION** Prior to usage, check the centrifuge tubes and their lids for damage. The polypropylene and polyethylene can withstand the toluene for a limited time, but eventually the material will break. If you see even small cracks or damaged spots, replace the centrifuge tubes.
>
> **! CAUTION** Make sure to thoroughly balance the centrifuge tubes. Two opposing centrifuge tubes should have a maximum weight difference of ±5 mg.

**24** Centrifuge the dispersion at 60,000 $g$ for 45 min. During the centrifugation step you have time to clean the Shear Force Mixer.

> ►**REFER TO Cleaning the Shear Force Mixer** (steps 30-34)

**25** Carefully remove the centrifuge tubes from the rotor and place them in a tube rack.

> ▲**CRITICAL** Try to move the tubes as little as possible, as any movement might stir up the pelletized, unexfoliated material.

**26** Using an Eppendorf pipette, take off the supernatant and fill it into a second set of centrifuge tubes. The supernatant should show a at least slightly purple colour. Again, balance out the centrifuge tubes.

> ▲**CRITICAL** This step will have a great impact on the dispersion quality! While taking up the supernatant, work slowly and stay close to the surface with the tip of the Eppendorf pipette. This will prevent any stir up of the pelletized material. Rather leave a little bit of supernatant in the centrifuge tubes than risking to transfer the pelletized material into the new set of centrifuge tubes.
>
> **! CAUTION** Prior to usage, check the centrifuge tubes and their lids for damage. The polypropylene and polyethylene can withstand the toluene for a limited time, but eventually the material will break. If you see even small cracks or damaged spots, replace the centrifuge tubes.
>
> **! CAUTION** Make sure to thoroughly balance the centrifuge tubes. Two opposing centrifuge tubes should have a maximum weight difference of ±5 mg.
>
> **? TROUBLESHOOTING**

**27** The pelletized material can be reused for subsequent dispersion steps.

> ►**REFER TO Recycling** (steps 36-41).

28  Centrifuge the supernatant of the first centrifugation step again at 60,000 *g* for another 45 min. This should pelletize any remaining unexfoliated material

29  Carefully remove the centrifuge tubes from the rotor and place them in a tube rack.
▲CRITICAL Try to move the tubes as little as possible, as any movement might stir up the pelletized, unexfoliated material.

30  Use a syringe to take off the supernatant and fill it into a storage container. For nicely dispersed nanotubes, you can use a syringe filter to remove any remaining particles. This is, however, not necessary if you worked carefully and can decrease the yield, if the dispersed nanotubes already started to aggregate.
▲CRITICAL Despite being polymer-wrapped, exfoliated SWNTs will start to reaggregate eventually; especially when dispersions have a high nanotube concentration. Before using them, sonicate the dispersion for 10 minutes, using a bath sonicator to break up any formed bundles. As reported by Schneider *et al.*, you can also use 1,10-phenantroline to stabilize your dispersions.[35]
■ PAUSE POINT The characterization of the dispersion can be delayed if you do not want to immediately recycle your pellets. Otherwise, you have to at least measure an absorption spectrum.
? TROUBLESHOOTING

**Cleaning the Shear Force Mixer**                                    ● **TIMING** 30 min

**! CAUTION** Wear gloves when handling parts of the Shear Force Mixer that were in direct contact with the nanotube dispersion

31      After the SFM has stopped, remove the metal aglet that connects rotor and stator by pushing it out with a suitable rod-shaped object. The rotor should now fall out. Disassemble the rest of the SFM. You will notice black stains on the parts of stator and rotor that were immersed in the nanotube dispersion. Try to remove them with IPA-wetted wipes. This works very well for the rotor and the outside of the stator.
**? TROUBLESHOOTING**

32      Clean the inside of the stator with IPA and a test tube brush of suitable thickness.

33      Immerse the stator in a measuring cylinder and fill it with toluene until all nanotube stains are covered. Place the measuring cylinder in an ultrasonication bath and sonicate for 15 mins

34      Remove the stator from the measuring cylinder and discard the dirty toluene. Rinse both, stator and measuring cylinder, with toluene.

35      After letting stator and rotor dry, reassemble the Shear Force Mixer as shown in **Figure 7**.
▲**CRITICAL** If any IPA is left and gets into the next dispersion, it will cause severe aggregation problems.

**Recycling of pelletized material**                                  ● **TIMING** 30 min

▲**CRITICAL** Running the shear force mixing procedure once does not exfoliate all the (6,5) SWNTs contained in the CoMoCAT® raw material. The pellets will still contain lots of (6,5) SWNTs and can usually be recycled 3-4 times before you have to use fresh SWNT material.

▲**CRITICAL** For recycling of the pelletized material, you have two options. You can either directly recycle the pellets after centrifugation (see steps 36-38) or you can let them dry (steps 39-41)

36  After step 26, a pellet of unexfoliated material, that still contains (6,5) SWNTs and a few millilitres of supernatant are left in the centrifuge tubes. Pour the remaining supernatant back into the Erlenmeyer flask. With toluene from a squeeze bottle, loosen the pellet and pour it back into the Erlenmeyer flask.

37  The pellet will contain some polymer but new polymer should be added. The extinction coefficient of PFO-BPy for the band at 363 nm is $96.4 \pm 0.2$ mL(mg cm)$^{-1}$ (see **Figure S3**). An absorption spectrum of the dispersion you obtained from the process that yielded the pellets should be measured.
  ►**REFER TO Characterization** (step 42)

38  The amount of to-be-added polymer is now calculated with:
$$m_{Polymer} = \frac{A_{363\,nm} \cdot V_{dispersion}}{\varepsilon_{363\,nm} \cdot d_{cuvette}}$$
  ▲**CRITICAL** As PFO-BPy is very selective towards (6,5) SWNTs and small variations of concentration won't influence your selectivity, you can roughly estimate $V_{dispersion}$.

38  Add the calculated amount of PFO-BPy to the Erlenmeyer flask that contains the pellets. Fill it with toluene until you have a total volume of ~200 mL in the flask. From here on, follow steps 14-35.

39  Instead of directly recycling the pelletized material, the pellets can be dried for use at a later point. The remaining supernatant should be removed and the centrifuge tubes with the pellets should be left open for drying for ~3 days.

40  Remove the dried pellets from the centrifuge tubes with a spatula and put them in a container. For longer storage times, store them in a dry environment.
  ■ **PAUSE POINT** You can store the pellets for several months before reusing them.

| 41 | To recycle the dried pellets, calculate the amount of polymer you need to add as explained in step 37 and 38. Add the dried pellets, the polymer and ~200 mL of toluene to an Erlenmeyer flask. Follow steps 14-35 from here. |
|---|---|

**Characterization of dispersions** ● **TIMING** 2 h

▲**CRITICAL** UV-vis-NIR absorption and resonant Raman spectroscopy are used to check whether the selective dispersion of (6,5) SWNTs was successful without dispersing any other semiconducting or metallic nanotube species.

| 42 | Measure an absorption spectrum of the dispersion in the range between 300-1600 nm. It should look as shown in **Figure 10** with SWNT-related peaks at 1000 nm ($E_{11}$), ~900 nm (phonon side band, PSB), 575 nm ($E_{22}$) and a peak at 363 nm originating from PFO-BPy.<br>▲**CRITICAL** To obtain a useful value for the polymer peak at 360 nm, you might have to dilute the dispersion. It should have an absorption of less than 2 cm$^{-1}$ in order to calculate the amount of polymer you need to add during recycling.<br>**? TROUBLESHOOTING** |
|---|---|

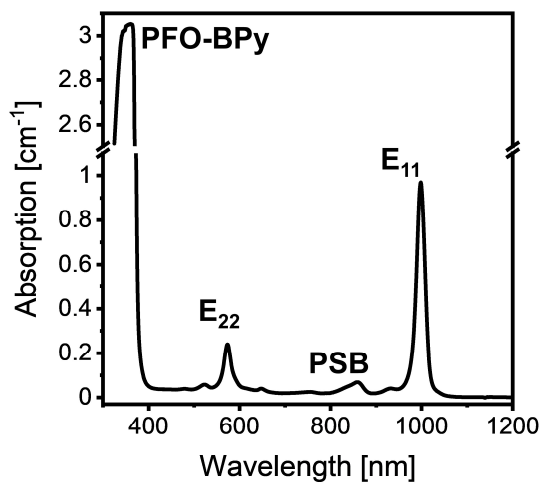

**Figure 10.** Absorption spectrum of a monochiral (6,5) SWNT dispersion obtained by Shear Force Mixing. The dominant peaks ($E_{11}$, PSB, $E_{22}$, PFO-BPy) are marked in the spectrum.

**43** For Raman measurements, drop-cast the dispersion onto a coverslip coated with aluminum foil to form a film (**Note**: aluminum-foil shows no Raman signal at all). The film should have a visible purple colour to obtain sufficient signal without too much integration. To avoid a background from excess PFO-BPy, you can remove it by carefully rinsing the film with toluene or THF. To check for other SWNT chiralities, especially metallic nanotubes, the region between 200-320 cm$^{-1}$, in which SWNTs show the diameter dependent radial breathing mode, has to be investigated. (6,5) SWNTs can be efficiently excited with a 532 or 785 nm laser. To account for spot-to-spot variations in the drop-cast sample, we measure at least 200 spectra at different positions (square grid) and average them. **Figure 11a** shows a spectrum of a drop-cast dispersion that only contains (6,5) SWNTs, while **Figure 11b** was measured on a dispersion that also contained metallic SWNT chiralities.

**? TROUBLESHOOTING**

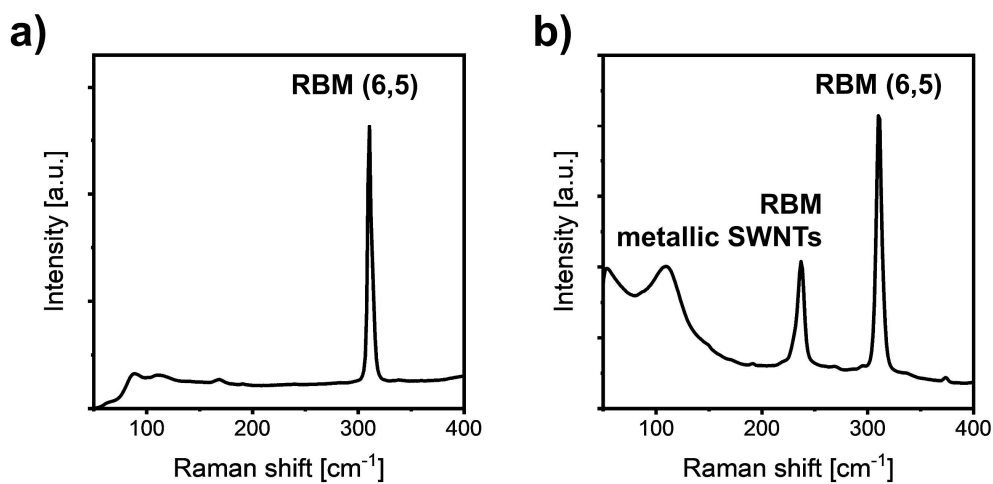

**Figure 11. a)** Raman spectrum of a monochiral (6,5) SWNT dispersion without contamination of metallic nanotubes as indicated by the absence of peaks aside from the RBM at 310 cm$^{-1}$. **b)** Raman spectrum of a (6,5) SWNT dispersion with a large content of metallic tubes. Metallic tubes lead to a peak at ~240 cm$^{-1}$ (usually not as pronounced as in this case). Both samples were excited with a 532 nm laser.

# TROUBLESHOOTING

## TROUBLESHOOTING Step 19

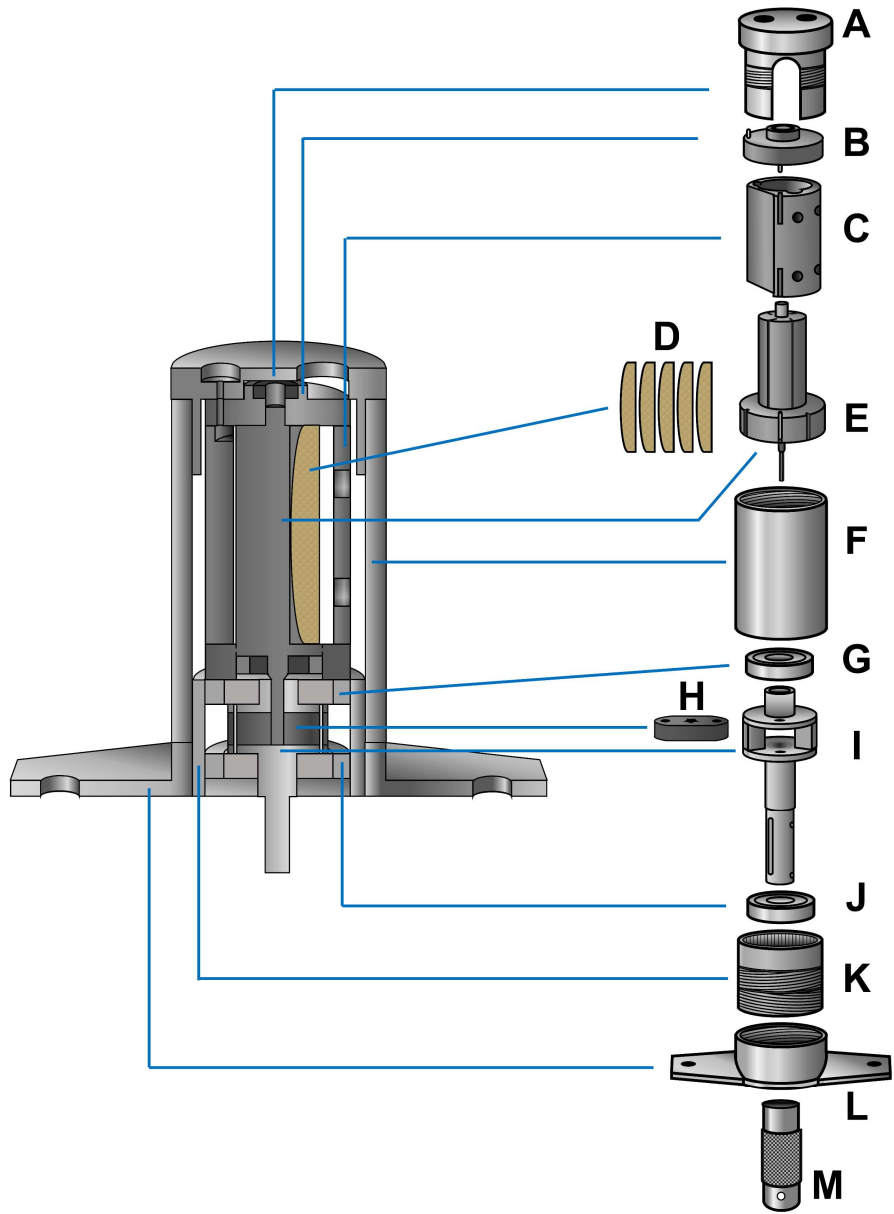

**Figure 12.** Cross section of the assembled motor unit and explosion drawing. In the explosion drawing all the parts are in the correct order for assembly. The motor unit consists of lid (B), stator (C), vanes (D) and rotor (E). The rounded site of the vanes is oriented towards the centre of the rotor. The rotor is connected to a shaft (I) *via* an aglet (H). To ensure smooth rotation of the shaft, two identical ball bearings (G and J) hold it inside the housing that consists of a lid (A), a housing (F) and a bottom part (K, L). Mounting the adapter (M) to the shaft (I) allows for connection to the rotor of the Shear Force Mixer. Note that all threads in the motor are left-handed threads.

- **?** **The pitch of the noise that the motor produces is not uniform.**

  **Most likely origin:** The noise is produced by moving parts inside the motor. The air-driven motor of the shear force mixer has five vanes (**D**), that grind down against the motor housing (**C**). Over time, a significant amount of debris builds up inside of the SFM (**Figure 13**). This debris can cause irregular motion of the motor unit, which leads to a non-uniform pitch of the whining noise.

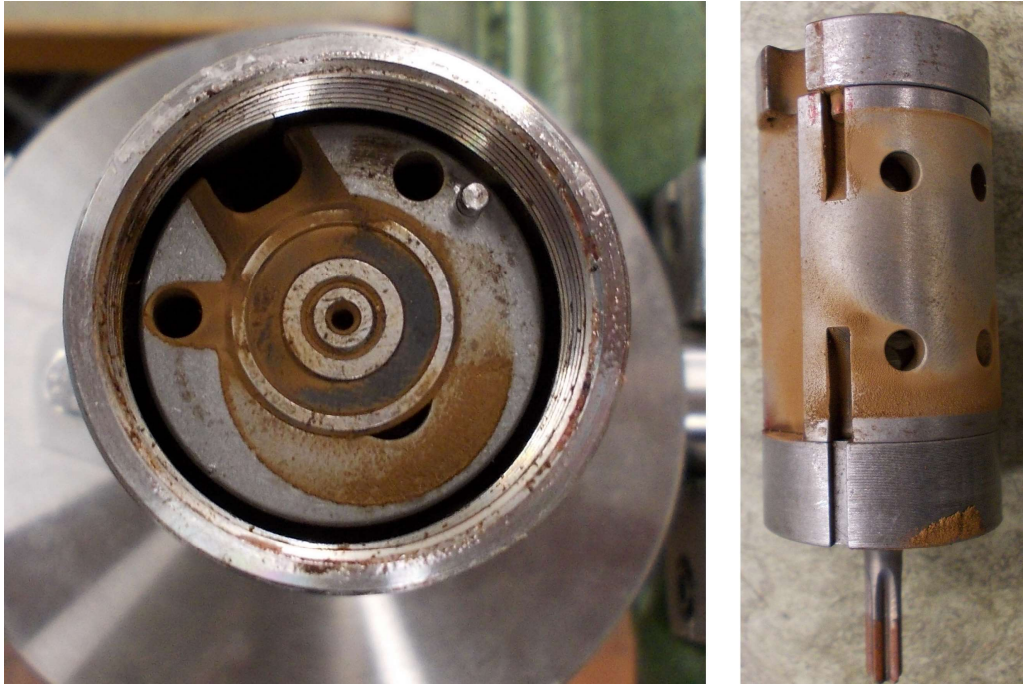

**Figure 13.** Motor with significant amount of debris (left: after unscrewing lid (A), right: only motor unit (B-E)). The debris can be removed with pressurized air and a damp towel.

**Solution:** The motor unit must be disassembled and cleaned. **Figure 12** might help to understand the how to disassemble (and later reassemble) the motor unit. Once disassembled, clean dirty parts with pressurized air from the orange-brown debris.

- **?** **The shear force mixer does not move at all or gets stuck regularly although you can hear air flushing through it.**

  **Most likely origin:** The air-driven motor of the shear force mixer has five vanes, that grind down against the motor housing. After several months of continuous usage, the vanes are significantly smaller than in the beginning (see **Figure 14**) and can fall out of their respective slits which will prevent rotation.

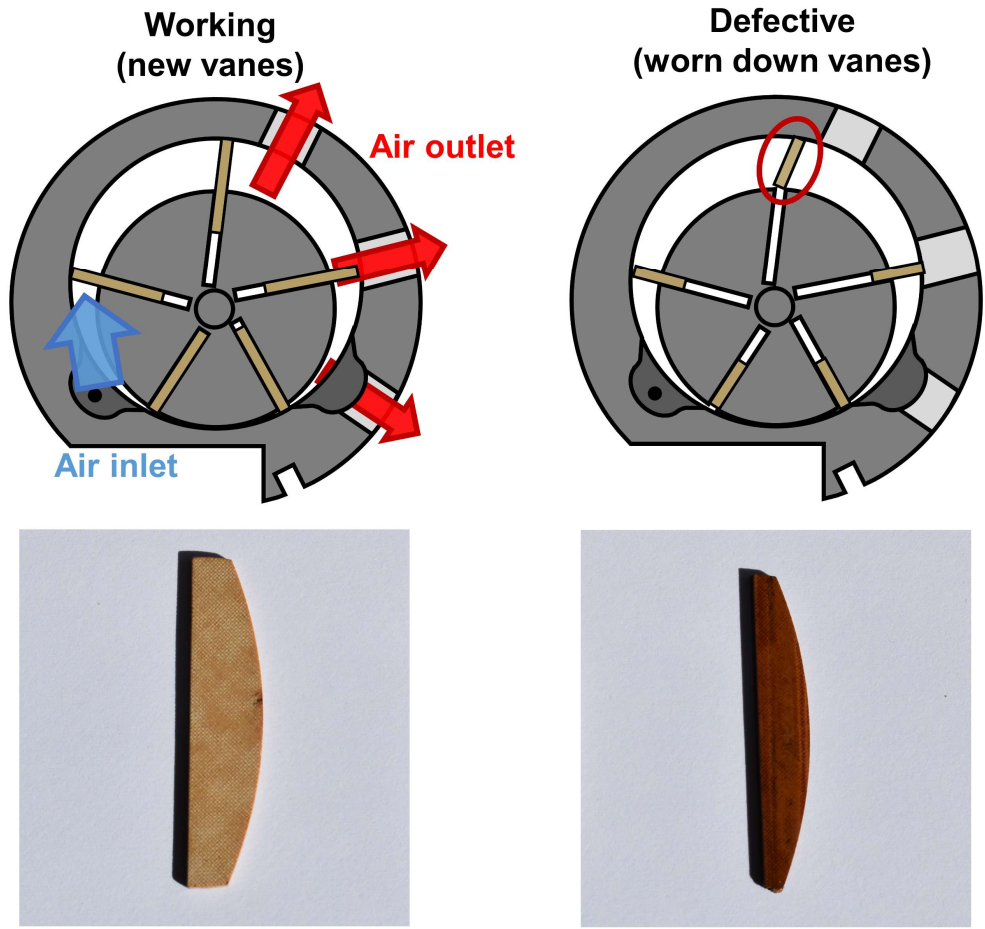

**Figure 14.** The top panel shows a schematic depiction of the working principle of an air-driven vane motor with new vanes (left) and significantly worn-down vanes (right) that can cause problems during operation. The picture of the vanes (bottom panel) shows a drastic change in vane size and shape.

Additionally, the metal parts will wear down over time. Especially the connection between motor and rotor (Parts E and H/I in **Figure 12**) is prone to failure after intensive use.

**Solution:** When the vanes are worn-down, you have to exchange them. A **repair kit**, that includes 5 vanes, gaskets and ball bearings can be ordered from Atlas Copco. In case of worn-down metal parts, a new motor unit must be ordered from Silverson.

**?**     **There are "scratching" noises.**

**Most likely origin:** The rotor grinds against the stator.

**Solution:** The PTFE bush (**Figure 3b**, **D**) that stabilizes the rotor in the stator must be replaced every few months to prevent this issue from happening. The brushes can be ordered individually from Silverson.

**TROUBLESHOOTING Step 26**

**?**     **The supernatant does not show any colour.**

**Most likely origin:** There are no nanotubes dispersed or the dispersion was not stable. This can have several causes:
- High humidity during dispersion
- Too little polymer
- Temperature was not controlled sufficiently

**Solution:** Check that your chiller works sufficiently and that you regularly empty the dehumidifiers. In case of very humid conditions, even the dehumidifiers might fail. For possible recycling runs, consider adding more polymer.

**?**     **There still is unexfoliated, non-pelletized material floating in the dispersion.**

**Most likely origin:** The centrifugation process is not yet complete.

**Solution:** Try to centrifuge for another 15 min. Make sure that you move the centrifuge tubes as little as possible when removing them from the centrifuge.

**?**     **The supernatant is purple but also shows a greyish colour.**

**Most likely origin:** This indicates either a non-complete sedimentation (see above) or exfoliation of metallic SWNTs.

**Solution:** Reduce the amount of polymer in the next (recycling) run.

**TROUBLESHOOTING Step 30**

?     **When using a syringe filter, there is significant resistance when pushing the dispersion through it and/or the syringe filter turns purple.**

**Most likely origin:** The nanotube dispersion has started aggregating and nanotube aggregates clog the PTFE filter.

**Solution:** If you need to use a syringe filter, try sonicating the dispersion for 5 minutes and filter immediately after sonication.

**TROUBLESHOOTING Step 31**

?     **The rotor gets stuck when trying to remove it from the stator.**

**Most likely origin:** Debris from the vanes obstructs the rotor movement.

**Solution:** Remove the rotor together with the stator by removing the metal aglet and the conical screws/nuts that fixate the stator. The rotor can now be pushed out of the stator. Clean both rotor and stator extensively to prevent any future problems.

## TROUBLESHOOTING Step 42

**?** **There are additional peaks in the nIR-region of the absorption spectrum.**

**Most likely origin:** Additional semiconducting chiralities were dispersed. You will usually disperse very little other chiralities (*e.g.*, (7,5) SWNTs) which will be visible in Raman but should not appear in absorption spectra

**Solution:** Reduce the amount of polymer in the next SFM run.

**?** **Over a longer period of time (>3 years), a decrease in yield is observed.**

**Most likely origin:** During years of continuous usage, the rotor can grind against the stator and thus enlarges the distance between them. This, in turn, leads to a decrease in shear force and thus a lower dispersion efficiency. We saw a substantial decrease after ~4 years of heavy usage (see **Figure 15**)

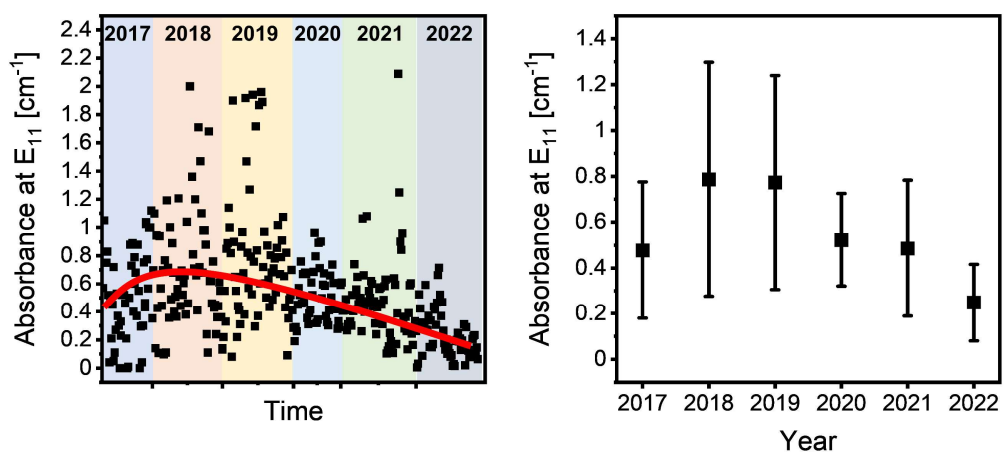

**Figure 15**. Evolution of the absorbance of (6,5) SWNT dispersions at the $E_{11}$ over several years in the Zaumseil group. The dispersions plotted here were obtained from the same shear force mixer. Left: all data points, right: averaged over the entire year.

**Solution:** A new rotor and shear head must be bought from Silverson. **Note**, you only have to replace the detachable shear head (E in **Figure 3**) and not the whole stator.

## TROUBLESHOOTING Step 43

? **There is only a broad background and no nanotube signals.**

**Most likely origin:** The sample has a high polymer-to-nanotube ratio.

**Solution:** Drop-cast a thicker film and carefully wash away excess amounts of polymer with THF.

? **Additional peaks between 200-250 cm$^{-1}$ appear**

**Most likely origin:** Metallic SWNTs have been exfoliated

**Solution:** Reduce the amount of polymer in the next SFM run.

? **There is an additional peak at ~280 cm$^{-1}$**

**Most likely origin:** (7,5) SWNTs have been exfoliated.

**Solution:** A small amount of (7,5) SWNTs is usually dispersed. A small peak is nothing to worry about. If the peak is substantial, reduce the amount of polymer in your next run

# SUPPORTING INFORMATION

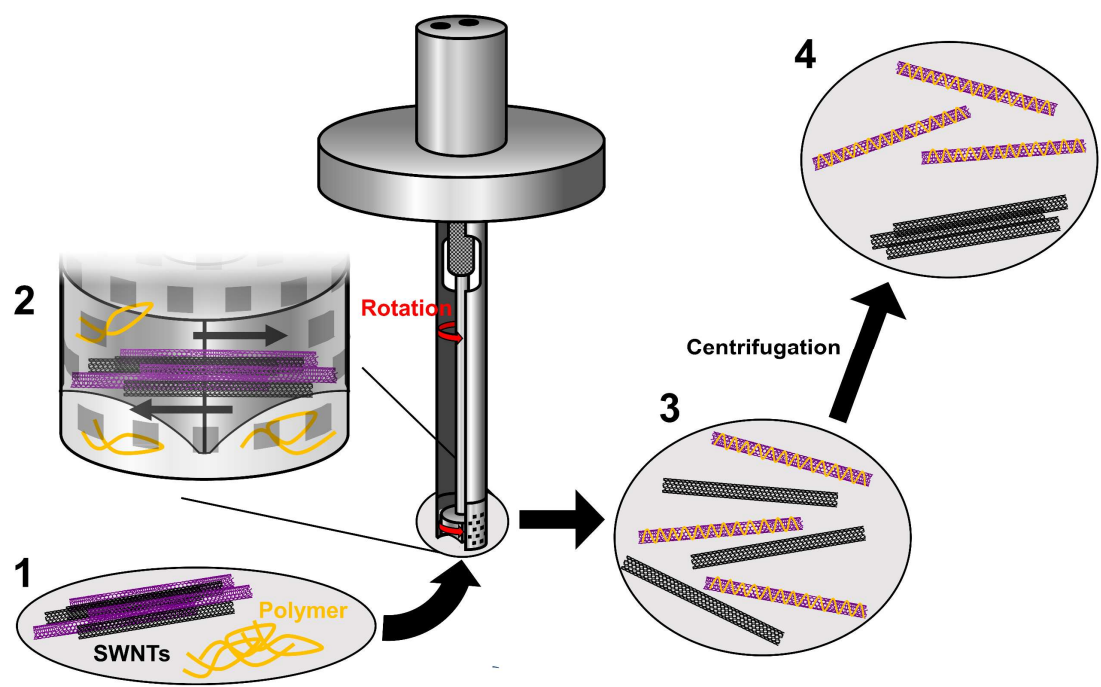

**Figure S1.** Working principle of a shear force mixer. During operation, the shear force mixer works like a pump. Solvent, unexfoliated SWNT raw material and polymer are sucked into the rotor/stator area where larger chunks are broken down (**1**). The fast rotation of the rotor produces centrifugal forces that push solids and solvent towards the edge of the shear screen. During this process the SWNTs are pushed out of the square holes in the screen, high shear forces act on the nanotube bundles and break them up (**2**). The polymer can now wrap around the exfoliated nanotubes but only stabilizes certain nanotube types, others re-aggregate instantly (**3**). These reaggregated SWNTs will be removed during the centrifugation step (**4**).

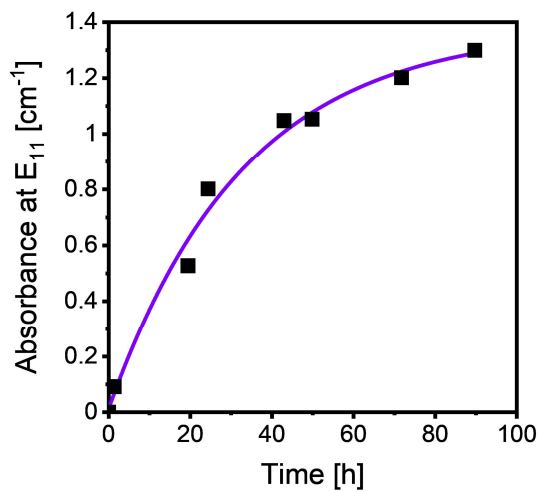

**Figure S2.** Relationship between duration of shear force mixing and dispersion yield. The data were obtained from Graf *et al.*[S1]

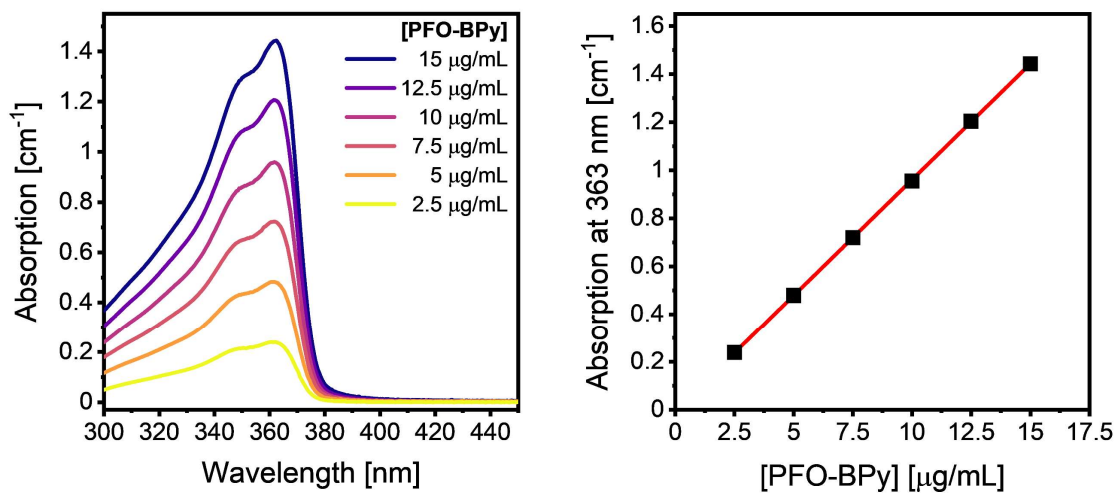

**Figure S3. a)** UV-Vis absorption spectra of PFO-BPy solutions in toluene. **b)** Lambert-Beer plot of the absorption (1 cm cuvette) at the absorption maximum at 363 nm against PFO-BPy concentration.

# APPENDIX I: REQUIRED CHANGES FOR DISPERSION OF (7,5) SWNTs

## Raw Materials

For (7,5) SWNTs a combination of CoMoCAT® raw material and poly[9,9-dioctylfluorene-2,7-diyl] (PFO) is used. The raw materials, that we use, are:

- **CoMoCAT® raw material** (Sigma-Aldrich, 773735, (6,5) chirality, ≥95% carbon basis (≥95% as carbon nanotubes), 0.78 nm average diameter)
- **poly[9,9-dioctylfluorenyl-2,7-diyl]** (PFO, Sigma-Aldrich, $M_W > 20$ kg mol$^{-1}$)

## Polymer Concentration

Precise control over the polymer concentration is necessary in the dispersion process of (7,5) SWNTs to obtain monochiral dispersions, as shown by the absorption data in **Figure S4a**.

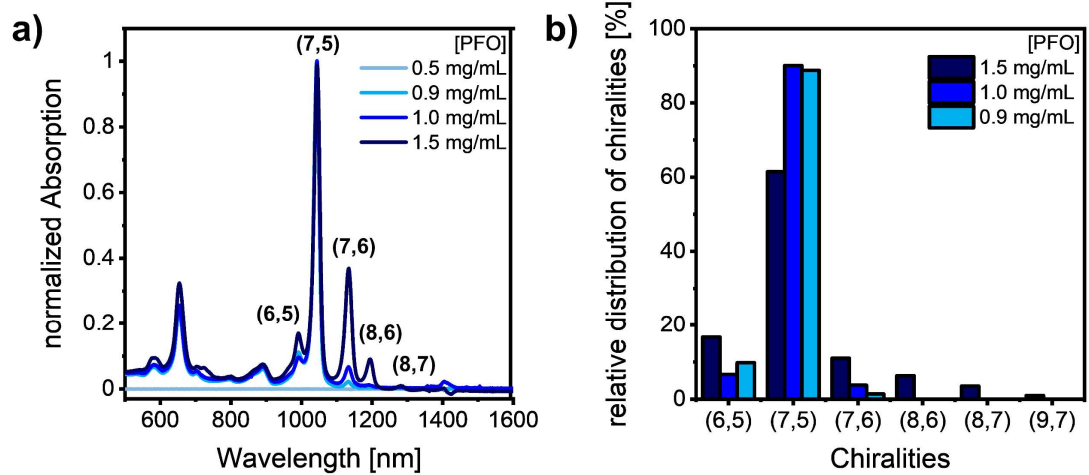

**Figure S4.** a) Absorption spectra of nanotube dispersions obtained by shear force mixing of CoMoCAT® raw material with varying PFO concentrations. b) Relative distribution of nanotube chiralities for varying PFO concentrations calculated with the script by Pfohl *et al.*[S2] [DOI: 10.1021/acsomega.6b00468] and experimentally determined extinction coefficients.

While insufficient polymer concentration (0.5 mg/mL) prevents any dispersion of nanotubes, a very high concentration (1.5 mg/mL) leads to dispersion of additional nanotube chiralities. Using the script by Pfohl et al.[S2] and experimentally determined extinction coefficients,[S3] which vary between nanotube chiralities, the relative chirality distribution can be calculated (see **Figure S4b**). The highest chiral purity (>90 %) is achieved for PFO concentrations of 0.9 and 1.0 mg/mL. Raman spectroscopy, however, reveals that for 1.0 mg/mL some metallic nanotubes are still exfoliated (see **Figure S5**).

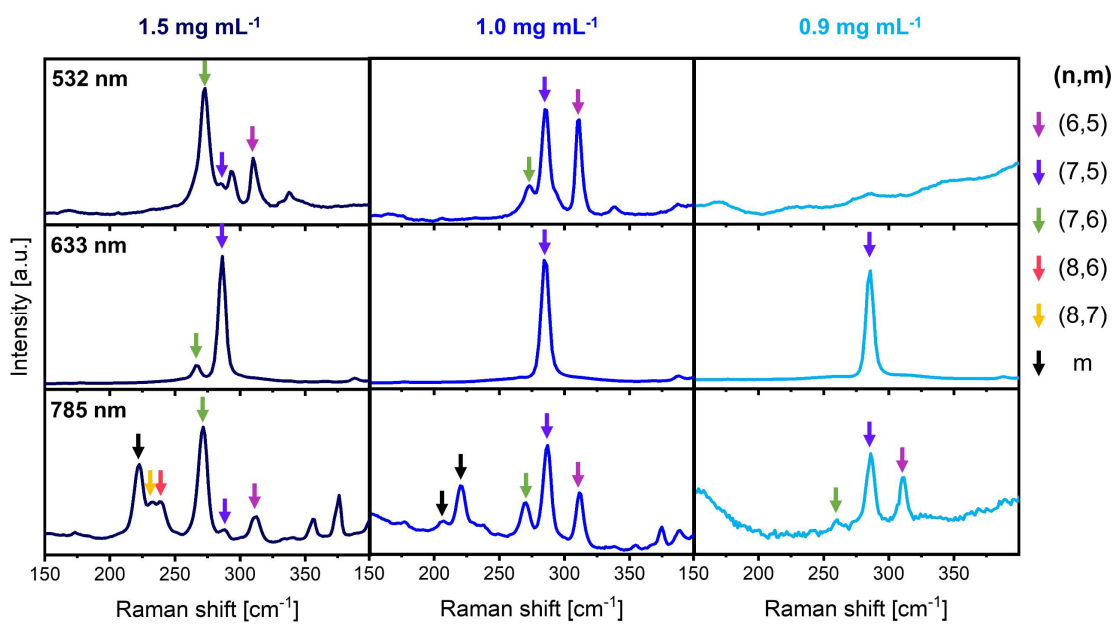

**Figure S5.** Raman spectra of drop-cast dispersions of CoMoCAT® nanotubes exfoliated with different PFO concentrations (1.5, 1.0 and 0.9 mg/mL) in toluene. Three different excitation lasers (532 nm, 633 nm and 785 nm) were employed to reveal as many chiralities as possible. RBMs of identified chiralities are marked with colored arrows.

Despite the reduced yield, we recommend using 0.9 mg/mL instead of 1.0 mg/mL as the PFO concentration to obtain electronic-grade dispersions with a high chiral purity.

## Recycling

As the polymer concentration is such an important parameter for dispersions of (7,5) SWNTs, one has to ensure the addition of the right amount of PFO for recycling. The extinction coefficient of PFO was determined to be 96.1 ± 0.2 mL(mg cm)$^{-1}$ (see **Figure S6**). The amount of polymer to be added is calculated in the same way as described in the protocol for (6,5) SWNTs.

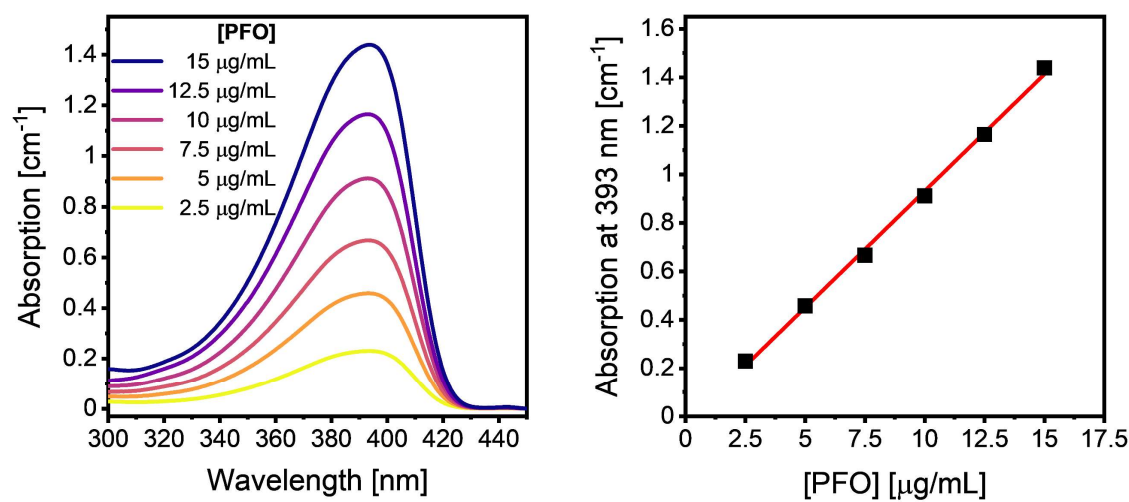

**Figure S6. a)** UV-vis absorption spectra of PFO solutions in toluene. **b)** Lambert-Beer plot of the absorption (1 cm cuvette) at the absorption maximum at 393 nm against PFO concentration.

## Characterization

The CoMoCAT® raw material contains significantly less (7,5) than (6,5) SWNTs. Thus, the overall dispersion yield will be lower than for dispersions for (6,5) SWNTs (usually the absorption at the $E_{11}$ peak is below 0.2 cm$^{-1}$). To check for other nanotube chiralities or metallic nanotubes in (7,5) SWNT dispersions with Raman spectroscopy, we advise to use at least a 633 nm and a 785 nm laser.

# APPENDIX II: REQUIRED CHANGES FOR DISPERSION OF PLASMA TORCH SWNTs

## Raw Materials

For the dispersion of only semiconducting Plasma Torch SWNTs, a combination of Plasma Torch nanotubes and PFO-BPy is used:

- **Plasma Torch SWNT raw material** (Raymor Industries Inc., RN-220, diameter 0.9-1.5 nm, batch RNB739-220-161220-A329)
- **Poly[(9,9-dioctylfluorenyl-2,7-diyl)-alt-(6,6′-[2,2′-bipyridine])]** (American Dye Source, PFO-BPy, ADS153UV, $M_w$ ~ 34 000 g/mol)

## SWNT concentration

As the Plasma Torch raw material has a lower SWNT content compared to the CoMoCAT® raw material, we use 1.5 mg/mL instead of 0.5 mg/mL of raw material. The polymer concentration remains unaltered at 0.5 mg/mL.

## Purification

The higher nanotube concentration will lead to bigger pellets, that tend to break apart more easily. Pay extra attention when removing the supernatant after the first centrifugation step. Additionally, Plasma Torch SWNTs tend to reaggregate quickly. Work fast once you stopped shear force mixing, to prevent sedimentation during the centrifugation. Do not use a syringe filter as it will get clogged almost immediately and the yield is decreased significantly.

## Characterization

Absorption spectra of Plasma Torch SWNT dispersions recorded between 400-1600 nm should show an $E_{33}$-region between 400-600 nm, an $E_{22}$-region between 800-1000 nm and the onset of an $E_{11}$-region at 1500 nm (**Figure S7**). Additional peaks between 600-800 nm are an indicator for metallic nanotubes. The residual background in this wavelength range should be as low as possible.

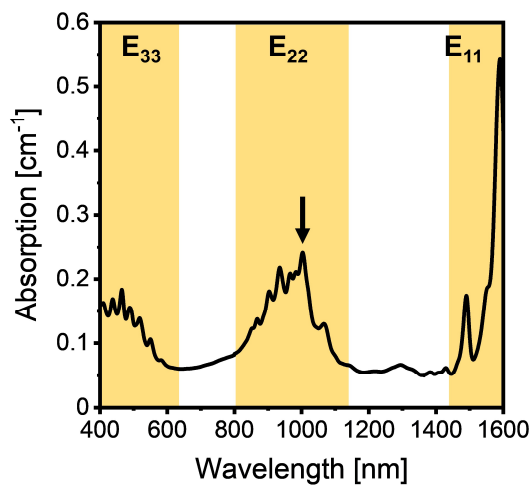

**Figure S7.** Absorption spectrum of a Plasma Torch SWNT dispersion with PFO-BPy. The absence of peaks in the range of 600-800 nm indicates a high semiconducting purity.

The arrow in **Figure S7** indicates the highest $E_{22}$ transition. The absorption value at this transition multiplied by 10 can be used to roughly compare the concentration of the obtained Plasma Torch dispersions to (6,5) and (7,5) SWNT dispersions.

Raman spectra of Plasma Torch SWNT dispersions must be measured at multiple excitation wavelengths (532 nm, 633 nm and 785 nm) to check for metallic nanotubes Raman spectra of a drop-cast Plasma Torch SWNT dispersion are shown in **Figure S8**. Regions, where RBMs of metallic nanotubes are expected are shaded in grey and marked by '**m**'.

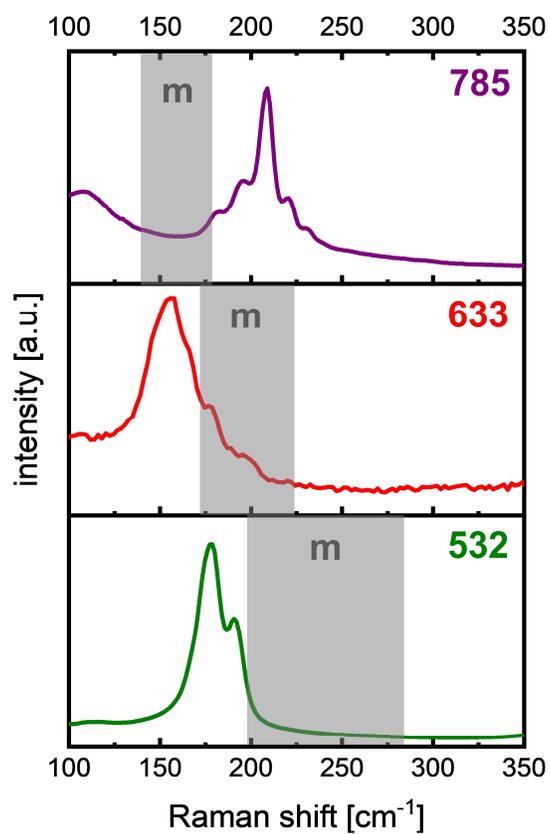

**Figure S8.** Raman spectra of drop-cast dispersions of semiconducting Plasma Torch SWNTs. Three different lasers (532 nm, 633 nm and 785 nm) were used for excitation to reveal as many chiralities as possible. Grey-shaded areas indicate a region where RBMs of metallic nanotubes would be expected for excitation with the respective laser.

# APPENDIX III: REMOVAL OF EXCESS POLYMER

High polymer concentrations (≥ 0.5 mg/mL) are necessary to selectively disperse nanotubes and help to stabilize the resulting dispersions. In processes such as $sp^3$-functionalization of SWNT dispersions or if clean SWNT films are desired, these polymer concentrations can lead to problems. Thus, we want to briefly describe how excess polymer can be removed *via* filtration (for polymer concentrations up to 1.0 mg/mL) or ultracentrifugation (for polymer concentrations >1.0 mg/mL). Note that removal of polymer will lead to very instable dispersions, that will aggregate quickly. In these cases you can add 1,10-phenantroline to your dispersions to stabilize them.[S4]

## Method A: Filtration

### Equipment:

- **PTFE filters** (Millipore, Omnipore Membrane Filter JHWP02500, hydrophilic PTFE Filter, 25 mm diameter, 0.1 µm pore size)
- **Glass vial** (4 mL volume)
- **Glass beaker** (250 mL volume)
- **Hotplate**
- **Ultrasonication bath** (here Branson 2510)
- **Membrane pump** (here Millipore vacuum pump WP6122050)
- **Glassware for the filtration setup as shown in Figure S9a**

### Chemicals:
- **Toluene** (purity >99.7 %)

### Procedure:

Fill toluene into the glass beaker and heat it to ~80°C on the hotplate. Assemble your filtration setup (**Figure S9b**). The filter should be centred above the metal mesh (**Figure S9c**). Test the setup for leaks by filtering a few millilitres of pure toluene through the PTFE membrane. If nothing leaks, fill the glass reservoir with ~25 mL of nanotube dispersion and filter it off. After this, wash the filter with ~10 mL of hot toluene. Repeat the process until you have filtered off the desired amount of nanotubes. The filtration process will get slower with the amount of nanotubes that you have already filtered off. Thus, consider filtering off large amounts of nanotubes onto several filters.

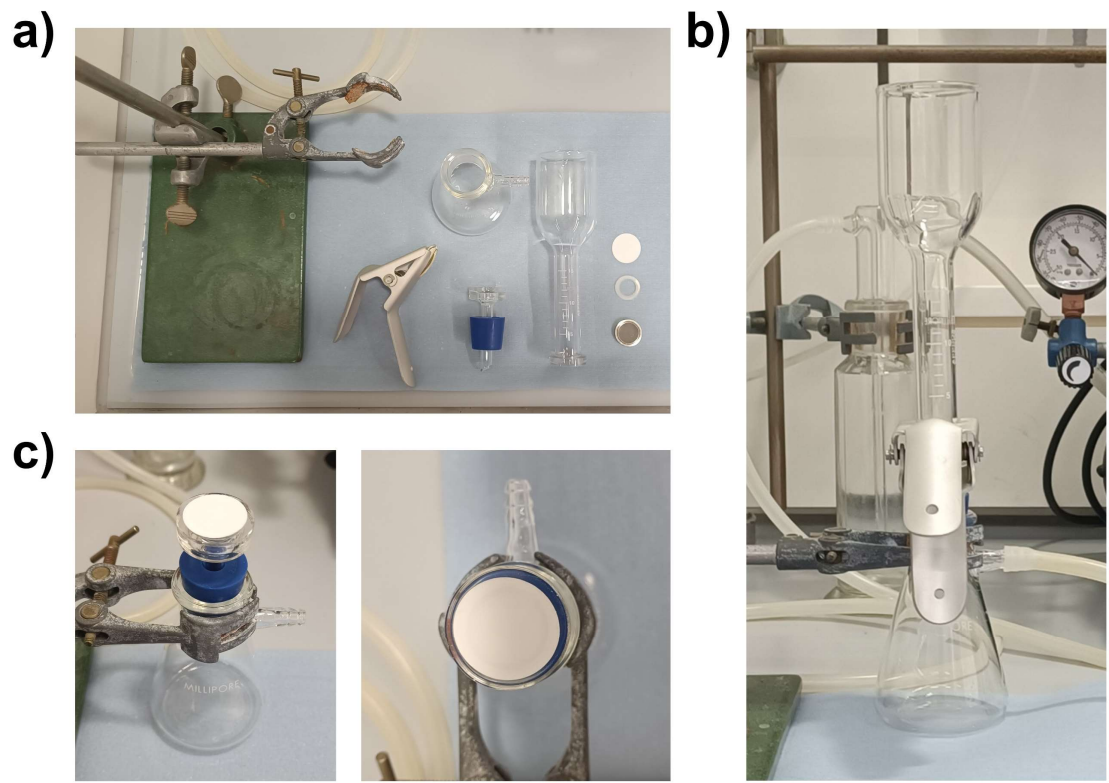

**Figure S9. a)** Disassembled filtration setup. **b)** Assembled filtration setup. **c)** Picture of ideal PTFE filter placement on the metal mesh.

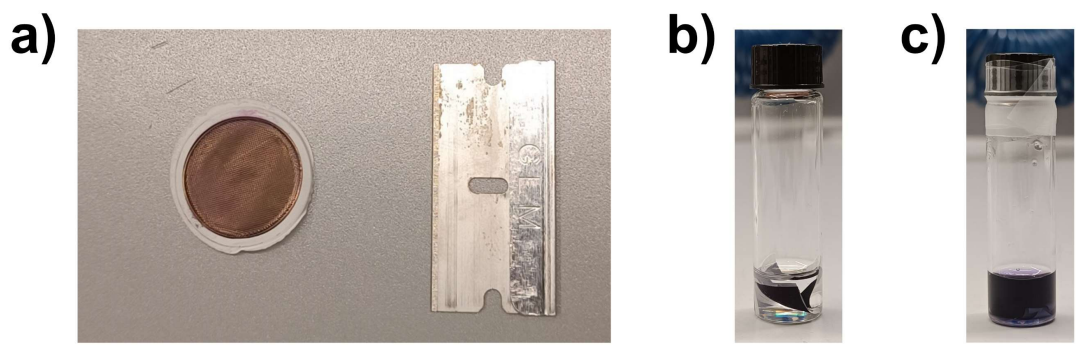

**Figure S10. a)** Filter containing filtered off (6,5)-SWNTs and razor blade used for quartering the filter. **b)** Quartered filter submerged in toluene before sonication. **c)** Dark purple dispersion of (6,5)-SWNTs in toluene obtained by sonicating the vial from Figure S10b for 20 min.

Carefully remove the clamp and the glass reservoir and take off the filter. Quarter the filter using a clean razor blade (**Figure S10a**). Transfer the quarters into a 4 mL vial and add 1 – 1.5 mL of solvent. Ensure, that the filter is immersed in solvent (**Figure S10b**). Wrap the lid with parafilm to prevent water from entering the dispersion. Sonicate the vial for 20 min in the ultrasonication bath to obtain a concentrated, polymer-free dispersion of SWNTs (**Figure S10c**).

## Method B: Ultracentrifugation

### Equipment:

- **Ultracentrifuge** (needs to reach RCF values of at least 250000 $g$, here Optima XPN-80, Beckmann Coulter)
- **Centrifuge Tubes** (need to have acceptable resistance against toluene and THF, here 15 mL Polypropylene Centrifuge Tubes, Beckmann Coulter)
- **Ultrasonication bath** (here Branson 2510)

### Chemicals:
- **Toluene** (purity >99.7 %)

### Procedure:

For polymer concentrations above 1 mg/mL, PTFE filters will clog very quickly. In these cases, we use ultracentrifugation to pelletize dispersed SWNTs.

Depending on your ultracentrifuge, the sample preparation (including type of centrifuge tubes, filling level, etc.) might vary. We centrifuge our samples at 284000 $g$ for 12 h using a swinging-bucket rotor (SW 40 Ti, Beckmann Coulter). The swinging-bucket rotor ensures an ideal pellet shape, which makes an unwanted stir-up less probable. After removing the centrifuge tubes from the buckets, carefully take off the supernatant and wash the pellet with THF and toluene. After washing, you can re-disperse the pelletized SWNTs in fresh solvent using an ultrasonication bath.

# APPENDIX IV: TECHNICAL DRAWINGS FOR SOUNDPROOF BOX K in FIGURE 4

The soundproof box consists of an inner box (1), an outer box (2) and a door (3). The inner box is made from polyoxymethylene plates and rests on four rubber feet inside the outer box (made from plywood). The space between the boxes is filled with sound dampening rock wool.

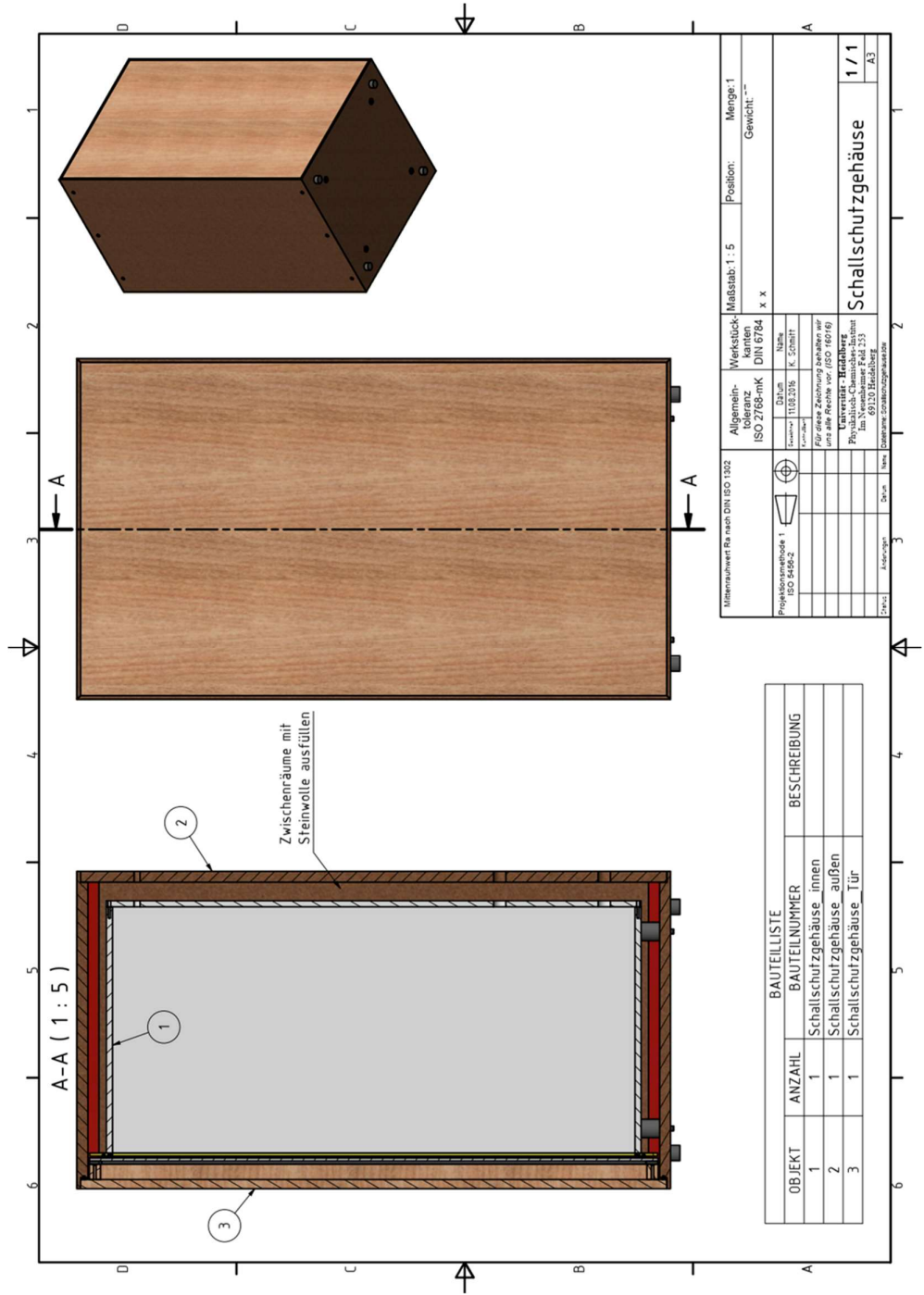

# REFERENCES SUPPORTING INFORMATION